\documentclass[sigconf]{acmart}

\usepackage[linesnumbered,ruled,vlined]{algorithm2e}
\usepackage{algpseudocode}

\usepackage{amsthm,amssymb,amsmath,graphics,float,subfigure}
\usepackage{empheq}
\usepackage{balance}
\usepackage{times}
\usepackage{bm}
\usepackage{diagbox}
\usepackage{multirow}
\usepackage{makecell}
\newcommand{\stitle}[1]{\vspace{1ex} \noindent{\bf #1}}
\long\def\comment#1{}
\newcommand{\kw}[1]{{\ensuremath{\mathsf{#1}}}\xspace}

\newcommand{\dbl} {\kw{DBLP}}
\newcommand{\ama}{\kw{Amazon}}

\newcommand{\epi}{\kw{Epinion}}
\newcommand{\cai}{\kw{Caida}}
\newcommand{\wik}{\kw{WikiV}}
\newcommand{\ski}{\kw{Skitter}}
\newcommand{\pok}{\kw{Pokec}}
\newcommand{\ork}{\kw{Orkut}}
\newcommand{\you}{\kw{Youtube}}
\newcommand{\gow}{\kw{Gowalla}}
\newcommand{\ber}{\kw{Berkstan}}

\newcommand{\fri}{\kw{Friend}}

\newcommand{\kcdsp}{\text{$k$-\kw{DSS}}\xspace}

\newcommand{\kcl}{{\ensuremath{\mathsf{KClist}}++}\xspace}
\newcommand{\sctl}{\ensuremath{\mathsf{SCTL}}\xspace}
\newcommand{\kclexact}{{\ensuremath{\mathsf{KEXACT}}}\xspace}

\newcommand{\ibatch}{\ensuremath{\mathsf{PBUpdate}}\xspace}
\newcommand{\psctl}{\ensuremath{\mathsf{PSCTL}}\xspace}

\newcommand{\ssctl}{\kw{SCTSample}}
\newcommand{\skcl}{\kw{KCLSample}}
\newcommand{\spath}{\kw{CPSample}}

\newcommand{\ccpath}{\kw{CCPATH}}

\def\kdss{\text{$k$-\kw{DSS}}\xspace}

\def\gd{\kw{GD}}
\def\cp{\kw{CP}}
\def\CP{\kw{CP}}
\def\ecp{\kw{SCT\text{-}CP}}
\def\edp{\kw{SCT\text{-}DP}}
\def\elp{\kw{SCT\text{-}LP}}
\def\SCT{\kw{SCT}}
\def\FW{\kw{FW}}
\def\C{\mathcal{C}}
\def\obj{\kw{obj}}

\def\P{\mathbb{P}}

\newcommand{\cond}[1]{\text{\kw{C#1}}}

\begin{document}
	
	\title{Scalable $k$-clique Densest Subgraph Search}

\settopmatter{authorsperrow=5}

\author{Xiaowei Ye}
\affiliation{%
		\institution{Beijing Institute of Technology}
		\city{Beijing}
		\country{China}
	}
\email{yexiaowei@bit.edu.cn}

\author{Miao Qiao}
\affiliation{%
	\institution{The University of Auckland}
	\city{Auckland}
	\country{New Zealand}
}
\email{miao.qiao@auckland.ac.nz}

\author{Ronghua Li}
\affiliation{%
	\institution{Beijing Institute of Technology}
	\city{Beijing}
	\country{China}
}
\email{ lironghuabit@126.com}

\author{Qi Zhang}
\affiliation{%
	\institution{Beijing Institute of Technology}
	\city{Beijing}
	\country{China}
}
\email{ qizhangcs@bit.edu.cn}

\author{Guoren Wang}
\affiliation{%
	\institution{Beijing Institute of Technology}
	\city{Beijing}
	\country{China}
}
\email{ wanggrbit@gmail.com}

	\begin{abstract}
		In this paper, we present a collection of novel and scalable algorithms designed to tackle the challenges inherent in the $k$-clique densest subgraph problem (\kcdsp) within network analysis. We propose \psctl, a novel algorithm based on the Frank-Wolfe approach for addressing \kcdsp, effectively solving a distinct convex programming problem. \textcolor{black}{\psctl is able to  approximate \kcdsp with near optimal guarantees.}  The notable advantage of \psctl lies in its time complexity, which is independent of the count of $k$-cliques, resulting in remarkable efficiency in practical applications. Additionally, we present \spath, a sampling-based algorithm with the capability to handle networks on an unprecedented scale, reaching up to $1.8\times 10^9$ edges. By leveraging the \ccpath algorithm as a uniform $k$-clique sampler, \spath ensures the efficient processing of large-scale network data, accompanied by a detailed analysis of accuracy guarantees. Together, these contributions represent a significant advancement in the field of $k$-clique densest subgraph discovery. In experimental evaluations, our algorithms demonstrate orders of magnitude faster performance compared to the current state-of-the-art solutions.
	\end{abstract}
	
	\maketitle

	\section{Introduction} \label{sec:intro}
		\begin{figure*}
		\hspace{-0.8cm}\includegraphics[width=1.05\linewidth]{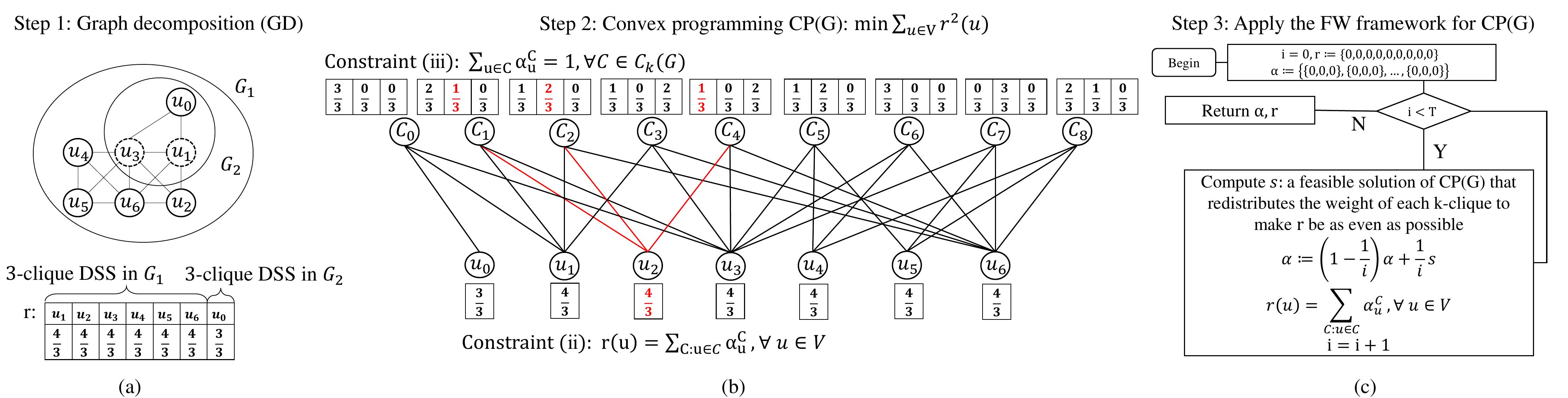}
			\vspace*{-0.7cm}
		\caption{The three-step paradigm for \kcdsp.}
		\label{fig:framework}
	\end{figure*}

	Dense subgraph search plays a primary role in graph mining. Given a graph $G$, a dense subgraph can be the subgraph with the highest edge-to-node ratio, called the densest subgraph, or the largest graph with all nodes mutually connected, called the maximum clique (whose size is denoted as $\omega(G)$). Both types of dense subgraphs can be captured by a unified problem called \emph{$k$-clique densest subgraph search} (\kdss)~\cite{kdensWWW2015}. Given an integer $k$ and graph $G$, \kdss reports a subgraph $S$ that maximizes the \emph{$k$-clique density} -- the ratio between the total number of $k$-cliques in $S$ and the number of nodes in $S$. \kdss reports the densest subgraph of $G$ when $k = 2$ and the maximum clique when $k = \omega(G)$. When $k$ is small, $\kdss$ reports dense subgraphs on small motifs such as triangles, which is applicable to document summarization~\cite{KS22} and network analysis (please see~\cite{LeeRJA10} as an entrance); when $k$ becomes larger, $\kdss$ reports near cliques (also called clique relaxations or defected cliques) where ``almost'' every pair of nodes are connected. Near clique search is important~\cite{MitzenmacherPPT15} to predicting protein-protein interactions~\cite{cui2008algorithm, DBLP:journals/bioinformatics/YuPTG06} and identifying over-represented motifs in DNA \cite{Fratkinbtl243}. 
	 Thus, the problem of \kdss has attracted an increasing attention recently \cite{kclpp,sctl}, exclusively on its computation efficiency and scalability.

	\begin{example}
		Figure~\ref{fig:framework}(a) shows a graph $G_1$ with $9$ triangles on $7$ nodes. Its $3$-clique density is thus $\frac{9}{7}$. The the subgraph induced by $\{u_1, u_2, \cdots, u_6\}$ achieves the maximum $3$-clique density of $\frac{4}{3}>\frac{9}{7}$.
	\end{example}
	

	
	\textcolor{black}{$\kdss$ is closely related to the problems of sampling, counting, or enumeration of $k$-cliques in a graph. It is worth noting that while approximate $\kdss$ can resort to sampling, exact $\kdss$  necessitates an enumeration of all  $k$-cliques \cite{kclpp,sctl, kdensWWW2015, MitzenmacherPPT15}. This complexity is lower bounded by the number of $k$-cliques on $G$, which limits the scalability. Having noticed this drawback, \sctl~\cite{sctl} propose to batch update by an index of $G$ called Succinct Clique Tree (\SCT)~\cite{PIVOTER}. $\SCT$ is a general technique that can disjointly partition all cliques of $G$ into groups. \sctl is based on the observation that some cliques in the same group are identical to update, and thus these cliques can be computed together in a batch. However, the availability of such a batch update is not guaranteed, and thus the enumeration of $k$-cliques may still be necessary in the worst case.}

	



	\begin{table*}[t]
			\vspace*{-0.3cm} 	
		\small
		
		\caption{Time and space complexities of SOTA approximate \kdss approaches of scanning-based (\kcl, \sctl) and sampling-based (\skcl,  \ssctl). $\delta$: the degeneracy of $G$ (shown in Table~\ref{tab:networks} for real dataset).  $\theta$: the arboricity of $G$ that $\delta/2 < \theta \leq \delta$. 
			$T$: the \# of iterations. $c_k = |\C_k(V)|$.  $c$: the \# of samples from $\C_k(V)$. $\eta$: the cardinality of \SCT-tree. $\kcl$ and $\sctl$ are dependent on $c_k$.}
	 \vspace*{-0.2cm} 	
		\label{tab:timememory}
		\centering
		\begin{tabular}{c | c | c |c || c | c |c }
			\toprule
			\textbf{SOTA Approx. }\kdss  & \kcl~\cite{kclpp} & \sctl~\cite{sctl} & \psctl (ours) & \skcl~\cite{kclpp}  & \ssctl~\cite{sctl} & \spath (ours) \\ 
			\midrule
			\textbf{Time}  $O(\cdot)$ & $km\theta^{k-2}+Tkc_k$ & $\theta^2 \eta + Tkc_k$ & $\theta^2 \eta + T\eta\theta^3$  & $km\theta^{k-2} + Tkc$ &  $\theta^2 \eta+Tkc$ &$n\delta^2k+(\delta k + k^2)t + Tkt$  \\
			\textbf{Memory}  $O(\cdot)$ &$kc_k$ & ${\theta\eta}$  & ${\theta\eta}$ &  $kc$ & $\theta\eta$ &  $kt$\\
			
			\bottomrule
		\end{tabular}
			
	\end{table*}

	This paper focuses on  \kdss \textbf{approximately} for better scalability. To find a \kdss solution whose complexity is \textbf{independent} of the number $c_k$ of $k$-cliques, we proposed to \textcolor{black}{relax \kcdsp into  a new convex programming problem $\ecp$.}  We prove that the optimal solution of $\ecp$ produces a \textcolor{black}{near-optimal approximation} of the $k$-clique densest subgraph. On all  datasets we tested in experiment, the optimal solution of $\ecp$ can find the optimal $k$-clique density exactly. More importantly, $\ecp$ allows one \textcolor{black}{to shift the dependency of the \kdss complexity on $c_k$ to the ``cardinality'' $\eta$ of the \SCT of $G$.  As  previous work \cite{PIVOTER} shows, $\eta$ is linear to the number $m$ of edges of $G$ in practice. On the real-world networks we tested in experiments, $\eta$ is less than $3m$, and is  $0.31m$ on average.} We compare the complexity of existing methods in Table~\ref{tab:timememory}, in which \psctl is the algorithm for finding the optimal solution of \ecp. To the best of our knowledge, this is the first approach whose complexity is independent of the number of $k$-cliques. Our solution achieves $k$-clique density comparable to the baselines while its running time is up to 4 orders of magnitude faster. 
	
	Sampling-based methods \textcolor{black}{for \kcdsp  \cite{MitzenmacherPPT15, kclpp, sctl} samples the $k$-cliques uniformly. The sampled $k$-cliques constructs a sparser graph \cite{MitzenmacherPPT15} and the $k$-clique densest subgraph in the sampled sparser graph is an approximation of the $k$-clique densest subgraph in the whole graph. Mitzenmacher et.al. \cite{MitzenmacherPPT15} proposed to sample a fixed proportion of the $k$-cliques to make sure the accuracy \cite{MitzenmacherPPT15}. However, the following works show that the sampling-based algorithms also have good performance when samples only a constant number of $k$-cliques, but there is no theory to explain why  \cite{kclpp, sctl}. Further,} 	sampling-based methods suffer from inefficient sampling process of the $k$-cliques. To sample $k$-cliques, existing algorithms either need to list all  $k$-cliques \cite{kclpp,MitzenmacherPPT15} or build the \SCT-index \cite{sctl}, leading to the exponential time complexity of sampling.  In this paper, we adopt $k$-color path sampling algorithm \ccpath~\cite{ccpath} as a uniform $k$-clique sampler, which has polynomial time complexity and dramatically improves the efficiency of sampling. The proposed algorithm is named as \spath. Notably, to the best of our knowledge, \spath is the first algorithm capable of effectively handling networks of up to $1.8\times 10^9$ edges in scale \textcolor{black}{for large $k$}. Additionally, we provide a comprehensive analysis of the accuracy guarantees. \textcolor{black}{Our analysis shows that the approximation has accuracy guarantees when the $k$-clique density of the subgraph reported by the sampling-based algorithms is large enough. The theoretical analysis is applicable to all existing sampling-based algorithms.}  
	In summary, we make the following contributions. 

	\begin{itemize}
		\item \stitle{\psctl: A Novel Frank-Wolfe Algorithm.} We formulate a new convex programming problem for \kdss approximation.  The new convex programming problem allows for weight assignment beyond the $k$-clique boundaries and is a near-optimal approximation for \kcdsp. We propose a novel Frank-Wolfe algorithm \psctl for solving the new convex programming problem. We prove that \psctl can achieve a near-optimal solution. Moreover,  the striking feature of \psctl is that the time complexity is independent of the count of $k$-cliques, making it highly efficient in practice.
		\item \stitle{\spath: Sampling-Based Algorithm for Large Networks.} We develop a scalable sampling-based algorithm, namely \spath. Specifically, \spath is an efficient algorithm that has polynomial time complexity and is capable of handling large networks. We also present a theoretical analysis \textcolor{black}{about the accuracy of  \spath, which is also applicable to existing sampling-based algorithms.}
		\item \stitle{ Extensive experiments.} We evaluate our algorithms on 12 large real-life graphs. The results show that \psctl is up to $4$ orders of magnitude faster than the state-of-the-art algorithms (\sctl and \kcl)  using similar space. Our sampling based algorithm \spath can obtain an good approximation and is also up to $4$ orders of magnitude faster than  the state-of-the-art sampling-based algorithms. For reproductivity purpose, we release our source code at \url{https://github.com/LightWant/densestSubgraph}.
	\end{itemize}


\section{Preliminaries}\label{sec:preliminaries}
	
	\begin{table}[t]
		\scriptsize
		\caption{Summary of notations}
		\label{tab:notations}
		\centering
			\begin{tabular}{c | c }
				\toprule
				\textbf{Notations} & \textbf{Descriptions} \\ 
				\midrule
				$G(V, E)$ & the graph $G$, node set $V$, edge set $E$ \\
				\hline
				$N(v, G)$ & the neighbors of $v$ in $G$ \\
					\hline
				$S$ & a subset $S\subseteq V$ \\
					\hline
				$\C_k(S)$ & the set of $k$-cliques in $G(S)$ \\
					\hline
				$P(V_h,V_p), V(P), \P$ & a SCT pair, $V(P)=V_h\cup V_p$, the set of all SCT pairs \\
					\hline
				$\delta$ & \makecell[c]{degeneracy of network} \\
					\hline
				$\eta$ & size of SCT \\
					\hline
				$\rho_k(S)$ & $k$-clique density of $G(S)$\\
					\hline
				$V^*$ & the set of nodes of the $k$-clique densest subgraph\\
					\hline
				$\alpha, \alpha^C_u$ (or $\alpha^P_u$) & \makecell[c]{$\alpha$ : the vector of weight assignment\\ from $k$-cliques (or SCT pair) to vertices;\\ $\alpha^C_u$  (or $\alpha^P_u$): the weight of the $k$-clique $C$ \\ (or SCT pair $P$) assigned to $u$} \\
					\hline
				$r(u)$ &  \makecell[c]{the rank of $u$ in the graph decomposition, \\ or the summary of weight assigned to $u$ \\ i.e. $r(u)=\sum_{C:u\in C}{\alpha^C_u}$}\\
					\hline
				$H(r)$ &\makecell[c]{ the first batch of $r$, which\\ is the set of nodes with the largest rank}\\
					\hline
				$t, T, p', p$ &\makecell[c]{$t$: sample size\\ $T$: number of iterations(a small constant) \\ $p'$: probability of a $k$-color path being a $k$-clique\\ $p=\frac{tp'}{|\C_k(V)|}$ : probability of $k$-clique being sampled}\\
				
				\bottomrule
			\end{tabular}
		\vspace*{-0.6cm}
	\end{table}
	
	Given an undirected graph $G(V, E)$ with node set $V$ and edge set $E$ where each edge $e(u,v) \in E$ has $u,v\in V$. Denote by $N(v,G) = \{u \in V|(u,v) \in E\}$ the neighbor set of $v$ in $G$. For a subset $S \subseteq V$ of nodes in $G$, we abuse $S$ to denote the induced subgraph of $S$ in $G$ whose node set is $S$ and edge set is $\{(u,v) \in E| u,v \in S\}$. $S$ is a clique for every pair $u$ and $v$ of nodes in $S$, $(u,v)\in E$. Consider a clique $C \subseteq V$ of $G$.  $C$ is a maximal clique if for any $v \in V\setminus C$, $C\cup \{v\}$ is not a clique. Denote by $\omega(G)$ the size of the maximum clique of $G$ and by $\C(G)$ the set of all  cliques in $G$. The degeneracy, denoted $\delta$, is the smallest value for which every subgraph $g$ of $G$ has a vertex with degree at most $\delta$ in $g$ \cite{13maximalclique, RongIO}. The value of degeneracy is often not very large in real-world networks \cite{13maximalclique, RongIO}. For an integer $k \in [2, \omega(G)]$, a clique $C$ is a $k$-clique  if $|C| = k$. $\C_k(S)$ denotes the set of $k$-cliques in the induced subgraph $S$. For a subgraph $S \subseteq V$, denote by $\C_k(S)$ the set of the $k$-cliques in $S$. The \textbf{$k$-clique density} of a subgraph $S$ of $G$ is 
	\begin{equation}\small
		\rho_k(S)=\frac{|\C_k(S)|}{|S|}.\nonumber
	\end{equation}
	
	\stitle{The \textit{$k$-clique Densest Subgraph Search} (\kcdsp).} Given  $G(V, E)$ and integer $k$. \kdss reports the $k$-clique densest subgraph of $G$.
	\begin{definition}[$k$-clique Densest Subgraph]
		The \textit{$k$-clique densest subgraph} of $G$ is a subgraph $V^*$ such that $\rho_k(V^*)\geq \rho_k(V')$ for all $V'\subseteq V$ and if $\rho_k(V^*) = \rho_k(V')$, $V'$ is a subgraph of $V^*$.  
	\end{definition}
	
	\stitle{Graph Decomposition (\gd)~\cite{kclpp, Danisch17}.} \textcolor{black}{$\gd$ is a $k$-clique density based decomposition of $G$. In $\gd$, iteratively ($i$-th iteration, $i\geq 1$) perform the following steps to produce the ranking $r(u)$ for each node $u\in V$. Given a graph $G$ on node set $V$ and integer $k$, let $V_1 = V$ and $\C^1$ be the collection of the $k$-cliques of $G$. For the next iterations, let $V_{i+1}$ be $V_i \setminus B_i$ and let $\C^{i+1}$ be $\{C | C\in \C^i, C\not\subseteq B_i\}$ where $\rho_i(S)=\frac{|\{C\in \C^i|C\subseteq S\}|}{|S|}$ and $B_i$ is the subgraph with maximum $\rho_i$ in $V_i$. Let $r(u)=\rho_i$ for all $u\in B_i$.  The iteration terminates when $V_i = \emptyset$. It was proved~\cite{kclpp} that for two iterations $i<j$, $\rho_i > \rho_j$. Clearly, $\rho_1(S) = \rho_k(S)$ and $B_1$ is the $k$-clique densest subgraph. Thus, the \kdss of $G$ equals to computing \textbf{the first batch} of $r$, i.e. $H(r)=B_1=\{u\in V| r(u) = max(r)\}$. }
	
	For example, for $k=3$, Figure~\ref{fig:framework}(a) shows the peeling process of $\gd$ where $S^* = \{u_1, \cdots, u_6\}$ with $\rho_1=\rho_k(S^*) = \frac{4}{3}$ in the first iteration. The vertices in $S^*$ have a rank value $r(u) =\rho_k(S^*) =\frac{4}{3}$.


	
	
	
	\stitle{Convex programming $\cp(G)$~\cite{kclpp,Danisch17}.} In the context of $\cp(G)$ below, three conditions (\cond{1\text{-}3}) assign a weight of 1 to each $k$-clique $C \in \C_k(V)$. This weight is then distributed among its nodes $u \in C$ using weights $\alpha_u^C \geq 0$ (\cond{1}, \cond{3}). Additionally, the total weights $r(u)$ accumulated by each node $u$ are subject to the condition that the squared summation is minimized (\cond{2}).  
	
	\begin{empheq}[box=\fbox]{align*}
		\cp(G) & & \min   \obj = \sum_{u \in V} r^2(u) && s.t.  \\
		\cond{1}&& \alpha^{C}_{u} \geq 0 & &  \forall  C \in \C_k(V), \forall u\in C \\
		\cond{2}&& \ r(u) = {{\sum_{C: u\in C \in \C_k(V)}{\alpha^{C}_{u}}}} && \forall u\in V\\
		\cond{3}&& \ \sum_{u\in C}{\alpha_u^C}  = 1 && \forall C\in \C_k(V)
	\end{empheq}
	
	It can be proved~\cite{Danisch17, kclpp} that the optimal solution of $\cp(G)$ is exactly the ranking $r(u)$ for the $\gd$ problem. Figure~\ref{fig:framework}(b) shows how the constraints $\cond{1}-\cond{3}$ facilitates the weight distribution in the optimal solution. The $3$-cliques on the top, $C_0$ to $C_8$ each has weight $1$ distributed to their 3 nodes linked to the second line ($u_0$ to $u_6$). The weights $r(u)$ collected by each node $u$ (shown in the box below the node) achieve the optimal squared sum in the objective function, which is identical to $r(u)$ computed in the $\gd$.

	\begin{lemma}[\cite{kclpp}]  \label{lemma:alpha0or1}
		Consider a $k$-clique $C$ and let $x$ be the node in $C$ with the smallest ranking and $y$ the node with the largest ranking. Either $r(y) = r(x)$ or $\alpha_y^C = 0$. 
	\end{lemma}
	
	\begin{lemma}[\cite{kclpp,Danisch17}] \label{lemma:CPequalGD}
		Let $V^*$ be the $k$-clique densest subgraph of $G$. Let $r(u), u\in V$ be the optimal solution of $\CP(G)$ and $S = H(r)$ the first batch of nodes in $r$. $S = V^*$ and $r(u) = \rho_k(V^*)$ for $\forall u\in S$.  
	\end{lemma}
	
	Apply Lemma~\ref{lemma:CPequalGD} recursively along the peeling process of the \gd can eventually reach the conclusion that the optimal solution of  $\CP(G)$ is exactly the ranking $r$ in $\gd$.

	\begin{algorithm}[t]
		\caption{The \FW Framework}
		\label{alg:framwork}
		\small
		\KwIn{The graph $G(V, E)$, clique size $k$, \# of iterations $T$}
		\KwOut{$\hat{V}^*$, an approximation of $V^*$}
		\SetKwProg{Fn}{Procedure}{}{}
		Initialize $r(u)$ for each $u\in V$\;
		Let $\C\C$ be either $\C_k(V)$ or a sample set of $\C_k(V)$\;
		\For{$t\gets 1,2,\cdots,T$}{
			\ForEach{$k$-clique $C\in \C\C$}{
				$u\gets \arg\min_{v\in C}{r(v)}$, break ties arbitrarily\;
				$r(u)\gets r(u)+1$\;
			}
		}	
		Sort $V$ by $r(u)$ in non-increasing order\;
		Denote $V_i$ by the previous $i$ vertices of the sorted $V$\;
		\Return{$\arg\max_{V_i} \rho_k(V_i)$}\;
		
	\end{algorithm}

	\stitle{Frank-Wolfe $(\FW)$ framework for solving \cp~(G).}  	Algorithm~\ref{alg:framwork} shows the \FW framework tailored for the convex programming $\CP(G)$ \cite{sctl,kclpp}. \FW iteratively (line~3) gets a clique $C$ in a random order (line~4) and then rebalances the weight distribution among all nodes in $C$ (lines~5-6).  Figure~\ref{fig:framework}(c) shows the flowchart of the \FW framework. Specifically, for \kdss, the "Compute $s$" step in $\FW$ is repeating \FW-1 and \FW-2 until $\C$ is empty. For the scanning based algorithm, $\C=\C_k(V)$. For the sampling based algorithm, $\C$ is the set of $k$-cliques uniformly  sampled from $\C_k(V)$.
\begin{enumerate}
	\item[\FW-1] Fetches a $k$-clique $C \in \C$, then removes $C$ from $\C$
	\item[\FW-2] Finds the node $x$ in $C$ that carries the lowest $r(x)$ (breaks ties arbitrarily), then redistributes the weight of $C$ to increase $r(x)$. 
\end{enumerate}

	\stitle{Remark.} All the SOTA methods, \kcl, \sctl, \skcl and \ssctl, follows the \FW framework of the convex programming \CP(G), and are different in the ways of selecting a clique $C$:  \skcl and \ssctl select $C$ based on a sample set $\C\C$ of $\C_k(V)$, and \kcl and \sctl scans $\C_k(V)$ either directly or using an index \cite{sctl, kclpp}. To remove the dependency of scanning-based method on $|\C_k(V)|$ while having the resulting density effectively bounded, we propose a new convex programming problem $\ecp(G)$.

	\stitle{Succinct Clique Tree~\cite{PIVOTER} (\SCT).} Given a graph $G(V,E)$, \SCT is an index of all cliques of $G$. It is produced by a set of large cliques of $G(V,E)$, where each large clique is presented as a pair $P=(V_h, V_p)$.   Each pair $P(V_h, V_p)$ consists of two disjoint node sets $V_h\subseteq V$ and $V_p \subseteq V$.  Denote by $V(P)=V_h\cup V_p$. The pair $P(V_h, V_p)$ encodes the set of cliques $\C(P)=\{V_h \cup C' | C'\subseteq V_p\}$. Since $P$ is a large clique, $V_h\cup C'$ for any $C'\subseteq V_p$ is a sub-clique encoded by $P$. Let $\P$ be the set of all pairs. Each clique is encoded by one and only one pair.  \SCT is an index of $\P$ in a special shape of  search tree.
	
	\begin{lemma}[\cite{PIVOTER}] \label{lemma:sctproperty}
		Let $P(V_h,V_p)$ be a pair in $\P$ of $\SCT$. We have $|\C(P)|=2^{|V_p|}$. In other words, the combination of any subset of $V_p$ and the entire $V_h$ forms a clique in $G$. In particular, $V(P)$ is a clique.  Besides, all cliques $\C(G)$ in $G$ are disjointly partitioned into $\{\C(P) | P \in \P\}$ \textcolor{black}{[Theorem.4.1., \cite{PIVOTER}]}. 
	\end{lemma}
	We call $\eta =|\mathbb{P}|$ the size of the $\SCT$ index which is a parameter related only to the graph $G$. The upper bound of $\eta$ is $O(n3^{\alpha/3})$ \cite{PIVOTER}. The value of $\eta$ is often not extremely large for real graphs \cite{PIVOTER} as shown in Table~\ref{tab:networks}. 
	
	\begin{lemma}
		Given a graph $G$ and its \SCT $\P$, an integer $k$, let $\P_k = \{P(V_h,V_p)\in \P|\ |V_h|\leq k\ \&\ |V(P)| \geq k\}$. For a pair $P(V_h,V_p) \in \P_k$, define all $k$-cliques in $\C(P)$ as set $\C_k(P) = \{V_h \cup C'| C'\subseteq V_p \& |C'| = k-|V_h|\}$. Then $\{\C_k(P)|P\in \P_k\}$ is a disjoint partition of all $k$-cliques $\C_k(V)$ of $G$. The cardinality $|\P_k|\leq |\P| = \eta$.
	\end{lemma}
	\begin{proof}
		\textcolor{black}{According to Lemma~\ref{lemma:sctproperty}, $\P$ covers all cliques. Thus, we can derive that $\P$ also covers all $k$-cliques. $\P_k$ comes from $\P$ by removing the pairs that $|V_h|>k$ and $|V(P)|<k$. When $|V_h|>k$, the size of the cliques $V_h\cup C'$ must be larger than $k$. When $|V(P)|<k$, the size of the cliques in $P$ must be smaller than $k$. $\P_k$ comes from $\P$ by removing the pairs that do not contain $k$-cliques.}
	\end{proof}

	In the following discussions, we focus on the $k$-cliques of $G$ and thus abuse $\P$ to denote $\P_k$ unless otherwised specified.  For easy reference, all the key notations in this paper are summarized in Table~\ref{tab:notations}.

	\begin{lemma}[\cite{PIVOTER}]\label{lem:vpcount}
		Given a pair $P(V_h, V_p)\in \P$, there are $|V_p|\choose k-|V_h|$ $k$-cliques in total. For each $u\in V_h$, there are $|V_p|\choose k-|V_h|$ $k$-cliques that contain $u$. For each $u\in V_p$, there are $|V_p|-1\choose k-|V_h|-1$ $k$-cliques that contain $u$.
	\end{lemma}
		To form a $k$-clique, we need to choose $k-|V_h|$ vertices from $V_p$. With $u\in V_p$ being chosen, we need to select $k-|V_h|-1$ vertices from $|V_p|-1$ vertices. 

\begin{figure}
	\vspace*{-0.5cm}
	\begin{center}
		\begin{tabular}[t]{c}
			\subfigure[An example graph]{
				\label{sfig:example}
				\includegraphics[width=0.33\linewidth]{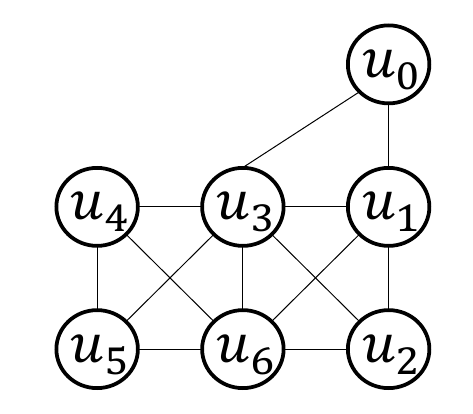}
			}
			\subfigure[An example of SCT-tree. In each path, the dotted nodes  are composed of the set $V_h$ and the solid  nodes  are composed of the set $V_p$.]{
				\label{sfig:example_sct}
				\includegraphics[width=0.62\linewidth]{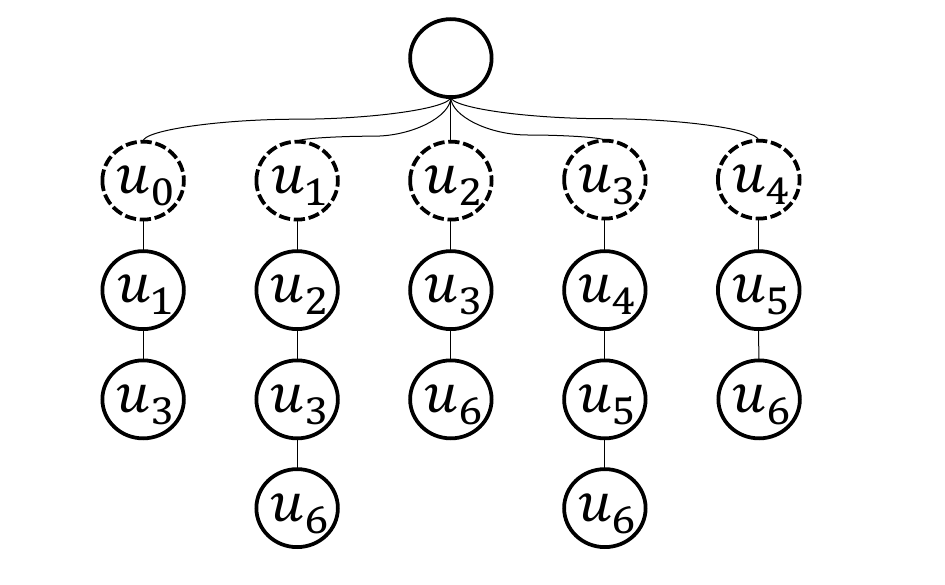}
			}
		\end{tabular}
	\end{center}
	\vspace*{-0.4cm}
	\caption{\color{black} Illustration of the SCT.}
	\vspace*{-0.2cm}
	\label{fig:example}
\end{figure}

\begin{example}
Figure~\ref{sfig:example_sct} is an example of SCT, where each path from root to leaf is a pair $(V_h, V_p)$. It has five pairs of $(V_h, V_p)$, including $(\{u_0\}, \{u_1,u_3\})$, $(\{u_1\},\{u_2, u_3, u_6\})$, $(\{u_2\}$, $\{u_3, u_6\})$, $ (\{u_3\}$, $\{u_4, u_5, u_6\})$, $(\{u_4\},\{u_5, u_6\})$. All the pairs are cliques, such as $G(\{u_3\}$, $\{u_4, u_5, u_6\})$ is a $4$-clique.  All the cliques in Figure~\ref{sfig:example} are uniquely encoded by the SCT. For example, the $3$-clique $\{u_3, u_5, u_6\}$ is encoded in the pair  $ (\{u_3\}, \{u_4, u_5, u_6\})$.
\end{example}

\section{New Convex Programming for $k$-DSS}\label{sec:psctl}
	To more  efficiently compute \kdss, we present a new convex programming problem {based on \SCT}. We first formulate $\ecp(G)$.

	\subsection{SCT-based Convex Programming}
	$\ecp(G)$ assigns a weight $\alpha_u^P>0$ for each node $u$ and each pair $P\in \P$ such that $u\in V(P)$. The assigned weights must satisfy the following constraints.


	\begin{empheq}[box=\fbox]{align*}
		\ecp(G)& \quad minimize \sum_{u\in V}{r(u)^2} \\ 
		s.t.\quad&  \cond{1}:  r(u) = \sum_{P\in \mathbb{P}\text{ s.t. } u\in V(P)}{\alpha^P_u}, \quad\quad\quad\text{ for }\forall u\in V\\ 
		&\cond{2}:  \alpha^P_u \ge 0 ,\quad \quad \quad\ \quad \quad \text{ for }\forall P \in \P \text{ and }  \forall u\in P\\ 
		&\cond{3}:  \sum_{ u\in V(P)}{\alpha^P_u} = {|V_p|\choose k -|V_h|}, \ \  \text{for }\forall P(V_h,V_p)\in \mathbb{P}\\  
		&\cond{4}: \alpha^P_u \le {|V_p|-1\choose k -|V_h|-1}, \forall P(V_h,V_p)\in \mathbb{P}, \forall u\in V_p
	\end{empheq}

	\stitle{Intuitive explanation.} Recall that in $\CP(G)$, each clique $C$ in $\C(G)$ is allowed to have a weight of $1$ distributed among the nodes $u$ in $C$, denoted as $\alpha_u^C$. Our new convex programming has made the following changes. Instead of each $k$-clique having weight $1$, we let each pair $P(V_h, V_p)$ have weight $|V_p|\choose k-|V_h|$ because there are $|V_p|\choose k-|V_h|$ $k$-cliques encoded by $P$ (Lemma~\ref{lem:vpcount}). We still want the weight to be retained in nodes in $C$ to avoid an over-flattened weight distribution in $V(P)$. To do so, we impose a soft constraint of $\cond{4}, \ecp(G)$ based on the fact that each node $u \in V_p$ participates in ${|V_p|-1 \choose k -|V_h|-1}$ $k$-cliques (containing $u$ and $V_h$), which gives an upper bound on $\alpha_u^P$.

\textcolor{black}{	Let $\alpha^*$ be the optimal solution of $\ecp(G)$.	Let $r^*$ be the ranking vector induced by $\alpha^*$. We first describe some properties of $\alpha^*$ and $r^*$. Then, we explain the relationship between $r^*$  and the $k$-clique densest subgraph.}

\stitle{\textcolor{black}{Properties of $\alpha^*$ and $r^*$.} }
Lemma~\ref{lem:propertyofr} shows that each entry $r^*(u)$ is upper bounded by the number of $k$-cliques that contain $u$. The property is identical to the optimal solution of $\cp(G)$ where each $k$-clique only assigns weight to one of the $k$ nodes in the $k$-clique.

\begin{lemma}\label{lem:propertyofr}
	Let $r$ be {a feasible} solution of $\ecp(G)$, it has $\sum_{u\in V} r(u) = |\C_k(V)|$ {and $\forall u\in V$, $r(u)$ is upper bounded by the number of $k$-cliques that contain $u$.}
\end{lemma}
\begin{proof}
	In accordance with $\cond{3}$ and $\ecp(G)$, all nodes in $V(P)$ share the weight of $|\C_k(P)| = {|V_p| \choose k-|V_p|}$ $k$-cliques. Since all $k$-cliques in $\C_k(V)$ are disjointly partitioned by \SCT (refer to Lemma~\ref{lemma:sctproperty}), it follows that $\sum_{u\in V} r(u) = |\C_k(V)|$. {Bynce \cond{3}, \cond{4} and Lemma~\ref{lem:vpcount}, we can achieve the upper bound of $r(u)$.}
\end{proof}

	\begin{theorem}\label{the:upper}
	 Given a pair $P=(V_h, V_p)$, define $up_v = {|V_p|\choose k-|V_p|}$ if $v\in V_h$ and $up_v = {|V_p|-1\choose k-|V_p|-1}$ if $v\in V_p$. ${\alpha^*}^P_v$ must equal  $up_v$  if there exists $u$ that $r^*(u) > r^*(v)$ and ${\alpha^*}^P_u>0$.
	\end{theorem}
	\begin{proof}
		Assume that $r^*(u) > r^*(v)$, ${\alpha^*}^P_u>0$ and ${\alpha^*}^P_v < up_v$. By re-assigning the weight from $u$ to $v$,  we can reduce the gap between $r(u)$ and $r(v)$ and reach a smaller value of the objective function, which contradicts  the fact that $\alpha^*$ is the optimal solution.
	\end{proof}

Theorem~\ref{the:upper} shows the property of $\alpha^*$ that the weight are assigned as even as possible. Since ${\alpha^*}^P_v$ is bounded by $up_v$  ($\cond{4}$ of $\ecp(G)$), $up_v$ is like a barrier to stop the weight flowing from the denser vertices to the $v$.

\textcolor{black}{We give the definition of \textit{relaxed stable subset}.}




\begin{definition}\label{def:relaxed}
	[\textbf{relaxed stable subset}]
	A subset of vertices $B$ is a relaxed stable subset if (1) $\forall u\in B, v\in V\setminus B, r(u) > r(v)$.
	(2) For each pair $P$ that intersects both $B$ and $V\setminus B$, $\exists u\in B\cap P, {\alpha^*}^P_u>0$ only when $\forall v\in P\cap (V\setminus B), {\alpha^*}^P_v=up_v$.
\end{definition}

\textcolor{black}{A relaxed stable subset $B$ has a larger rank than other nodes (condition~(1) in Definition~\ref{def:relaxed}) and the pairs that intersect both $B$ and $V\setminus B$ assign their weights to the nodes outside of $B$ as much as possible (condition~(2) in Definition~\ref{def:relaxed}).  }

\textcolor{black}{Recall that $H(r) =\{u\in V| r(u) = max(r)\}$. We can derive that $H(r^*)$ is a relaxed stable subset by Theorem~\ref{the:upper} and Definition~\ref{def:relaxed}. For a pair $P$ that intersects $H(r^*)$ and $V\setminus H(r^*)$, the $k$-cliques encoded by $P$ can be classified into three types: (1) all nodes in $H(r^*)$; (2) all nodes in $V\setminus H(r^*)$; (3) the nodes intersecting both $H(r^*)$ and $V\setminus H(r^*)$. Each of the $k$-clique has a weight of $1$. Since the weight of $P$ are assigned to nodes in $V\setminus H(r^*)$ as much as possible, the type~(1) $k$-cliques assign the weights to $H(r^*)$, and the type~(2) and (3) assign weights to $V\setminus H(r^*)$. In other words, the nodes in $H(r^*)$ only receive weights from the $k$-cliques in $H(r^*)$. }




	

\stitle{\textcolor{black}{Relationships between $r^*$ and $V^*$.}}
\textcolor{black}{ With the above analysis, we can derive Theorem~\ref{the:subset} and \ref{the:appGuarantee2} which explain the relationships between $r^*$ and the $k$-clique densest subgraph $V^*$.}

\begin{theorem}\label{the:subset}
\textcolor{black}{	The optimal solution of \ecp represents a subgraph $H(r^*)$ that $V^*\subseteq H(r^*)$.}
\end{theorem}
\begin{proof}\color{black}
	We prove that for any  $V'\subset V$ with $V'\cap (V\setminus \hat{V*}) \neq \emptyset$, $\rho_k(V')< \rho_k(H(r^*))$. By the analysis above, the nodes in $H(r^*)$ only receive weights from the $k$-cliques in $H(r^*)$. Thus,  we have 
	\begin{equation}\small
		\begin{aligned}
		\rho_k(H(r^*)) &\ge \frac{\sum_{u\in H(r^*)}{r(u)}}{|H((r^*))|}
	 \ge  \min_{u\in H(r^*)}{r(u)}
		\\ & > \max_{u\in V\setminus H(r^*)}{r(u)}  \ge  \rho_k(V').
		\end{aligned}		
	\end{equation}

\end{proof}

\begin{theorem}\label{the:appGuarantee2}
	
	
	

\textcolor{black}{The optimal solution $r^*$ of \ecp represents a subgraph $\hat{V^*}= H(r^*)$  that $\rho_k(\hat{V^*}) / \rho_k(V^*) >  1-\frac{ 1}{k|V^*|} $. 	}
\end{theorem}
\begin{proof}\color{black}
	Given two set of vertices $V_1$ and $V_2$, the set of additional $k$-cliques of $V_2$ to $V_1$ is $A(V_1,V_2)=\C_k(V_1\cup V_2) \setminus \C_k(V_1)$. From 
	\begin{equation}\small
		\rho_k(V^*) = \frac{|\C_k(V^*)|}{|V^*|} > \rho_k(\hat{V^*})=\frac{|\C_k(V^*)|+|A(V^*, \hat{V^*}\setminus V^*)|}{|V^*|+|\hat{V^*}\setminus V^*|},
		\nonumber
	\end{equation}
	we can obtain
	\begin{equation}\label{equ:additional1}\small
		|A(V^*, \hat{V^*}\setminus V^*)| <|\hat{V^*}\setminus V^*||\rho_k(V^*)|.
	\end{equation}
	
	Let $x$ be $|\C_k(V^*)|-|V^*|\rho_k({\hat{V^*}})$, which denotes the count of $k$-cliques in $\C_k(V^*)$ that assign their weight to $\hat{V^*}\setminus V^*$. According to the definition of SCT, each $k$-clique exists in only one pair. Thus, we can consider each pair independently. Given a pair $P=(V_h, V_p)$ that there are $x'_P$ $k$-cliques in $\C_k(V^*)\cap \C_k(P)$ assign their weight to  $V_p \cap (\hat{V^*}\setminus V^*)$, each vertex in $V_p \cap (\hat{V^*}\setminus V^*)$ generates at least $kx'_P$ additional $k$-cliques. Thus, we have at least $ k|V_p \cap (\hat{V^*}\setminus V^*)|x'_P $ additional $k$-cliques for pair $P$. Sum all the pairs together and we reach
	\begin{equation}\label{equ:additional2}\small
		|A(V^*, \hat{V^*}\setminus V^*)| \ge k\sum_{P(V_p,V_h)\in \P}{|V_p \cap (\hat{V^*}\setminus V^*)|x'_P}\ge kx|\hat{V^*}\setminus V^*|.
	\end{equation}

	Equation~(\ref{equ:additional2}) comes from the fact that $\sum_{P(V_p,V_h)\in \P}{x'_P}=x$ and $\sum_{P(V_p,V_h)\in \P}{|V_p \cap (\hat{V^*}\setminus V^*)|} \ge |\hat{V^*}\setminus V^*|$. The summary  is greater than the size of nodes outside of $V^*$ because the same node may be in more than one pairs.
	
	Combines Equation~(\ref{equ:additional1}) and (\ref{equ:additional2}) and we get $kx < \rho_k(V^*).$ Then, from the definition of $x$, we have
	
	$$\small k\left (|\C_k(V^*)|-|V^*|\rho_k({\hat{V^*}}) \right ) < \rho_k(V^*).$$ 
	
	At last, we can obtain the result
	\begin{equation}\small
		\frac{\rho_k({\hat{V^*}})}{\rho_k(V^*)} > 1-\frac{1}{k |V^*|}.
		\nonumber
	\end{equation}	
	
	
\end{proof}


Theorem~\ref{the:appGuarantee2} provides a guarantee that the optimal solution of our new convex programming is a near-optimal approximation of the $k$-clique densest subgraph.

\stitle{\textcolor{black}{The problem solved by $\ecp(G)$.}}	\textcolor{black}{Actually, $H(r^*)$ is the subgraph with the maximum value of $\rho'(S)=\frac{\sum_{P(V_h,V_p)\in \P(S)}{{|V_p|\choose k-|V_h|}}}{|S|}$, where $\P(S)$ is the set of pairs with all nodes in $S$. The nodes sorted by $r^*$ is a graph decomposition (\gd in Section~\ref{sec:preliminaries}) with respect to $\rho'(S)$. The statement is proved by  Theorem~\ref{the:ecptoelp} and Theorem~\ref{the:elpans}.}

\begin{theorem}\label{the:ecptoelp}
	The optimal solution $\alpha^*$ of $\ecp(G)$ is the optimal solution of $\edp(G)$.
	\begin{empheq}[box=\fbox]{align*}
		\edp(G)& \quad minimize \ \rho' \\ 
		s.t.\quad&  \cond{1}:  \rho' \ge r(u) = \sum_{P\in \mathbb{P}\text{ s.t. } u\in V(P)}{\alpha^P_u}, \quad\quad\quad\forall u\in V\\ 
		&\cond{2}:  \alpha^P_u \ge 0 ,\quad \quad \quad\ \quad \quad \quad \quad\forall P \in \P \text{ and }  \forall u\in P\\ 
		&\cond{3}:  \sum_{ u\in V(P)}{\alpha^P_u} = {|V_p|\choose k -|V_h|}, \ \  \quad \quad\forall P(V_h,V_p)\in \mathbb{P}\\  
		&\cond{4}: \alpha^P_u \le {|V_p|-1\choose k -|V_h|-1}, \forall P(V_h,V_p)\in \mathbb{P}, \forall u\in V_p
	\end{empheq}
\end{theorem}
\begin{proof}
Since the first level set $H(r^*)$ is a relaxed stable subset, all  weights of $H(r^*)$ come from the pairs in $H(r^*)$ and can not further distribute the weights to $V\setminus H(r)$. Therefore, $r^*(u)$ for $u\in H(r)$ is the minimum value that the objective function of $\edp(G)$ can achieve.
\end{proof}

	
	It is easy to derive that the dual of $\edp(G)$ is $\elp(G)$.
	\begin{empheq}[box=\fbox]{align*}
		\elp(G)& \quad \max \sum_{P(V_h,V_p)\in \P} {{|V_p|\choose k-|V_h|} x_P} \\ 
		s.t.\quad&  \cond{1}:  x_P \le y_u, \quad \quad \quad \forall u\in P\\ 
		&\cond{2}:  \sum_{u\in V} y_u = 1\\ 
		&\cond{3}:  y_u\ge 0, x_P\ge 0, \forall u\in P\in \P\\
	\end{empheq}

\begin{theorem}\label{the:elpans}
	Given a subgraph $S$, denote by $\P(S)$ the set of pairs $\{P| V(P)\in S, P\in \P\}$.
	The optimal solution of $\elp(G)$ is  the subgraph that maximizes $\rho'(S)=\frac{\sum_{P(V_h,V_p)\in \P(S)}{{|V_p|\choose k-|V_h|}}}{|S|}$.
\end{theorem}
\begin{proof}
	Given a subgraph $S$, let $y_u=\frac{1}{|S|}$ if $u\in S$ and $y_u = 0$ otherwise. Then, we can easily derive that the objective function of $\elp(G)$ is $\rho'(S)$. Thus, each subgraph corresponds to a feasible of $\elp(G)$.
	
	Given a feasible solution with value $\rho$, we now prove that it can also construct a subgraph $S$ that $\rho'(S)\ge\rho$. Given a parameter $z$, let $\P(z)=\{P|x_P\ge z\}$ and $S(z)=\{u| y_u\ge z\}$. According to $\cond{1}$ (\elp) and $P\in \P(z)$, we can derive that $u\in S(z), \forall u\in V(P), \forall P\in \P(z)$. Also, from $u\in S(z), \forall u\in V(P)$  we can derive that  $P\in \P(z)$. Suppose $\forall z, \frac{\sum_{P(V_h,V_p)\in \P(z)}{{|V_p|\choose k-|V_h|}}}{|S(z)|}< \rho$. Then, we have $\sum_{P(V_h,V_p)\in \P} {{|V_p|\choose k-|V_h|} x_P} < \rho,$ which is a contradiction. Therefore, there must exist a $z$ such that $\rho'(S(z))\ge \rho$.
	
Since the subgraphs and feasible solutions are mapped to each other, the subgraph with the maximum value $\rho^\prime$ maps to the optimal solution.
\end{proof}

From Theorem~\ref{the:ecptoelp} and Theorem~\ref{the:elpans}, we know that the optimal solution of $\ecp(G)$ finds the subgraph $S$ that maximize $\rho'(S)=\frac{\sum_{P(V_h,V_p)\in \P(S)}{{|V_p|\choose k-|V_h|}}}{|S|}$. Thus, the optimal solution of $\ecp(G)$ is intuitively a good approximation for the $k$-clique densest subgraph. As shown in our experiments, the optimal solution of $\ecp(G)$ on all  real-world datasets we tested can generate the $k$-clique densest subgraph exactly. 

%
%

\subsection{FW-based Algorithm for SCT-CP(G)}
	\begin{algorithm}[t]
		\caption{The Proposed \psctl Algorithm}
		\label{alg:psctl}
		\small
		\KwIn{Graph $G(V, E)$, clique size $k$, \# of iterations $T$, \SCT $\P$}
		\KwOut{$\hat{V}^*$, an approximation of $V^*$}
		\SetKwProg{Fn}{Procedure}{}{}
		Initialize $r(u)$ for each $u\in V$\;
		\For{$t\gets 1,2,\cdots,T$}{
			\ForEach{ pair $P\in \P$}{
				$r \gets \ibatch(r, P, k)$
			}
		}	
		Sort $V$ by $r(u)$\;
		Denote $V_i$ by the previous $i$ vertices of $V$\;
		\Return{$\arg\max_{V_i} \rho_k(V_i)$}\;
		
	\end{algorithm}
	
	\begin{algorithm}[t]
		\caption{\ibatch}
		\label{alg:ibatch}
		\small
		\KwIn{Ranking $r$, $P(V_h, V_p)$, clique size $k$, }
		\KwOut{Updated $r$}
		Partition $V(P)$ into $\{V_1, V_2,...,V_s\}$ based on a sorted ranking $r$.  Nodes in $V_i$ share the $i$-th smallest ranking $r_i$, $i\in [1,s]$, $r_1 < r_2 < \cdots < r_s$, $V_s = H(r)$\;
		$n_C\gets {|V_p|\choose k-|V_h|}$; \tcc{Total \# of cliques in $P$} 
		$p\gets |V_p|$, $h\gets |V_h|$\;
		\textbf{foreach} $u\in V_h$ $\textbf{do}$ $up(u) \gets {p \choose k-h}$\; 
		\textbf{foreach} $u\in V_p$ $\textbf{do}$ $up(u) \gets {p-1 \choose k-h-1}$\;
		$i\gets 1$\;
		\While{$n_C > 0$ and $s \geq i$} {\tcc{Allocate the weight of $n_C$ cliques to the nodes in $V_{i}$ to make $r$ be evenly}
			\textbf{if} $s \leq i$ \textbf{then} $gap \gets +\infty$ \textbf{else} $gap\gets r_{i+1}-r_{i}$\;
			\While{$n_C > 0$ and $\ gap > 0$ and $|V_{i}|>0$} {
				$w\gets \min\{\min_{u\in  V_{i}} up(u), gap, \lfloor\frac{n_C}{|V_{i}|}  \rfloor\}$\;
				\If{$w > 0$} {
					\For {$\forall u\in V_{i}$} {$r(u)\gets r(u)+w$; $up(u) \gets up(u)-w$\;}			
					$n_C \gets n_C - w\times |V_{i}|$; $gap\gets gap-w$\;
					
				}\Else{
					$r(u)\gets r(u)+1$ for a total of $n_C$  nodes $u$ in $V_{i}$ chosen uniformly at random\;
					$n_C\gets 0$\;
				}
				$V_{i}\gets V_{i} \setminus  \{u| up(u)=0 \& u\in V_{i}\}$\;
				
			}
			\textbf{if} $s \geq i + 1$ \textbf{then} $V_{i+1}\gets V_{i+1}\cup V_{i}$\;
			$i \gets i+1$\;
		}
		
		\Return $r$\;
		
	\end{algorithm}
	
	Algorithm~\ref{alg:psctl} tailors \FW framework for $\ecp(G)$. It replaces Lines~4-6 of Algorithm~\ref{alg:framwork} with  Algorithm~\ref{alg:ibatch} called for each $P\in \P$ to update the rankings $r$. 
	
	
	Algorithm~\ref{alg:ibatch} allocates the weight of $n_C$ (Line~2) cliques of pair $P$ to its nodes $V(P)$. The allocation aims at bringing up the rankings of the lowest-ranking nodes in $P$ so as to minimize the squared sum of $r(u), u\in V(P)$. This is done by first sorting all nodes in $P$ based on its original ranking and find the ``plateaus'' of different  rankings (Line~1): nodes in $V_s$ share the highest ranking $r_s$ (also called the first batch $H(r)$), nodes in $V_{s-1}$ the second largest ranking $r_2$, $\cdots$, nodes in $V_1$ the smallest ranking $r_1$. Denote by $s$ the number of different rankings of nodes in $V(P)$. The rankings $r_1$, $r_2$, $\cdots$, $r_s$ are strictly increasing. Lines~4-5 setup the upper bound for each node in the allocation based on $\cond{3}$ and $\cond{4}$ of $\ecp(G)$. 
	Lines~7-20 progressively use the budget of $n_C$ to bring the lowest ranking nodes up while satisfying the upper bounds set for each node. Specifically, $V_i$ (initially $i = 1$) is the set of smallest ranking nodes $u$ in $V(P)$ who can receive weights from the cliques without violating their upper bounds $up(u)$, we call them \emph{current potential nodes}. Denote by $gap$  the ranking differences of the smallest and the second smallest rankings of nodes $u$ in $V(P)$; if the second smallest ranking does not exist, then we let $gap$ be $+\infty$ (Line~8). After that, Lines 10-20 allocates the weights of $n_C$ cliques to the current potential nodes. In Line~10, $w$ the highest ranking increment of the potential nodes before i) a node reaches its upper bound and then should be kicked out of current potential node set $V_{i}$, ii) all  nodes in $V_i$ reaches the ranking of $r_{i+1}$, or iii) the weights of all $n_C$ cliques are used up. Each node in $V_{i}$ will then receive the weights of $w$ cliques (Lines~11-14) unless the $n_C < |V_i|$ (Lines~15-17) where a random set of $n_C$ nodes in $V_i$ will receive one extra weight.  $n_C$, $gap$, and the current potential set $V_i$ are updated accordingly (Lines~14,17,18). If by the time when $gap = 0$, there are still potential nodes in $V_{i}$, it means that their rankings have already reached $r_{i+1}$, we then merge them to $V_{i+1}$ (Line~19). 

	\begin{theorem}\label{the:psctltime}
		The time complexity of \psctl is $O(\eta \delta^3)$ where $\eta$ is the upper bound of the size of SCT-tree, and $\delta$ is the degeneracy of the input graph $G$. The space complexity is $O(\alpha \eta)$.
	\end{theorem}
	\begin{proof}
	In Algorithm~\ref{alg:ibatch}, the while loop in line~7, line~9 and line~12 all has time complexity $O(|V(P)|)$. Thus, the time complexity of Algorithm~\ref{alg:ibatch} is $O(|V(P)|^3)$. Since $P$ is a clique, $|V(P)|$ is bounded by $\delta$. \psctl need to scan over $\P$ and call \ibatch for each pair. Thus, the time complexity of \psctl is $O(\eta \delta^3)$. The main space cost of \psctl is the storage of the SCT-tree, which is $O(\alpha \eta)$ \cite{PIVOTER}.
	\end{proof}
	
	\begin{figure}[th]
				\vspace*{-0.5cm}
		\begin{center}
			
			\includegraphics[width=0.8\linewidth]{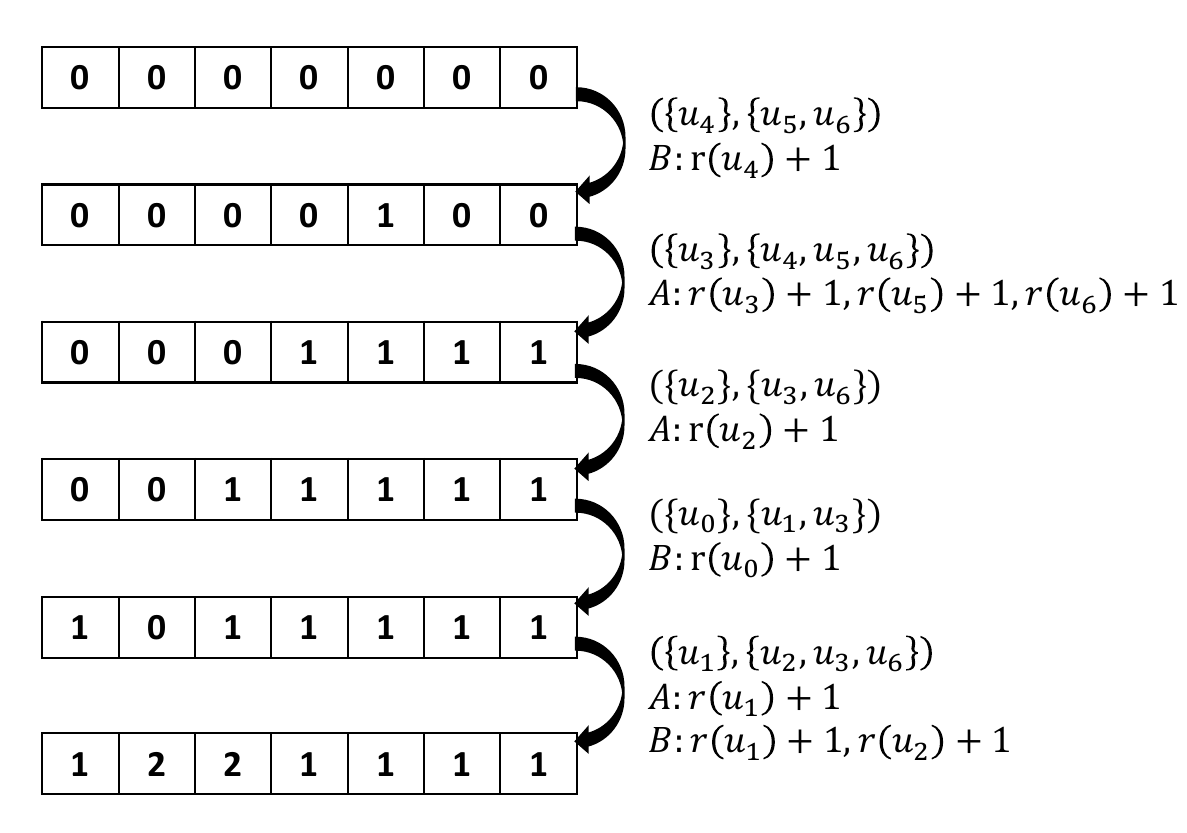}
		\end{center}
		\vspace*{-0.3cm}
		\caption{Illustration of \psctl on the example graph for one iteration.}
		\vspace*{-0.2cm}
		\label{fig:pupdate_example}
	\end{figure}
	
	\begin{example}\color{black}
		Figure~\ref{fig:pupdate_example} illustrates an example to explain how \psctl works on the example graph in Figure~\ref{sfig:example}.  In Figure~\ref{sfig:example}, it has five pairs of $(V_h, V_p)$, including $(\{u_0\}, \{u_1,u_3\})$, $(\{u_1\},\{u_2, u_3, u_6\})$, $(\{u_2\}$, $\{u_3, u_6\})$, $ (\{u_3\}, \{u_4, u_5, u_6\})$, $(\{u_4\},\{u_5, u_6\})$. The five pairs are accessing in a random order. Initially, the elements of the vector $r$ are all zeros. The updated label $A$ denotes that the update is line~13 of Algorithm~\ref{alg:ibatch}, and $B$ is line~16 of Algorithm~\ref{alg:ibatch}.
		
		Since there is only one $3$-clique in $(\{u_4\},\{u_5, u_6\})$ and the size of $V_{i}$ is $3$ (line~10 of Algorithm~\ref{alg:ibatch}), we have $\lfloor\frac{n_C}{|V_{i}|}  \rfloor=0$  and the weight is assigned to $u_4$ in 1ine~16 of Algorithm~\ref{alg:ibatch}. Then, for $ (\{u_3\},\{u_4, u_5, u_6\})$, there are $3$ $3$-cliques. Since $\{u_3,u_5,u_6\}$ has the smallest weight, the $3$ weights are assigned to $\{u_3,u_5,u_6\}$ averagely (line~13 of Algorithm~\ref{alg:ibatch}). The following updates are similar.  \hspace*{\fill}\mbox{$\Box$}
	\end{example}

	\stitle{Remarks.} Compared to the worst-case time complexity  $O(k|\C_k(V)|)$ of \sctl~\cite{sctl}, $O(\eta \delta^3)$ is much smaller for large real-world networks: as shown in Table~\ref{tab:networks}, the values of $\eta$ of \SCT-tree and degeneracy $\delta$ on large real-graphs are not large~\cite{PIVOTER}. More importantly, $O(\eta \delta^3)$ is independent of $k$ and the number of $k$-cliques in graph $G$.

	\subsection{Analysis of the Algorithm}
	Denote by $\alpha$ the vector of variables of $\alpha_u^P$ for $\forall P\in \P, u\in V(P)$. Let $\mathcal{D}$ be the domain of $\alpha$ specified in $\ecp(G)$. Since the rankings $r(u), r\in V$ are derived from $\alpha$, we denote the objective function of $\ecp(G)$ by \[f(\alpha) = \sum_{u\in V}r(u)^2.\]

	\begin{theorem}\label{the:ibatch_convex}
		\psctl is an implementation of a Frank-Wolfe algorithm to the convex program of $\ecp(G)$.
	\end{theorem}
	\begin{proof}
		\psctl solves the following problem.  Given $\alpha\in \mathcal{D}$, find $\hat \alpha \in \mathcal{D}$ to minimize $\langle \hat{\alpha}, \nabla f(\alpha) \rangle = 2\sum_{P\in \mathbb{P} : u\in P} {\hat{\alpha}^P_u \cdot r(u)}$. Since each pair has constant count of weights and can only assign weights to the vertices $V(P)$ in the pair (\cond{3}), we consider each pair independently. For each pair $P$, to minimize $\langle \hat{\alpha}, \nabla f(\alpha) \rangle$, the weights should be assigned to the vertices with the smallest value of $r$ in $\hat{\alpha}$. This is the work \ibatch do for each pair.
	\end{proof}

	By Theorem~\ref{the:ibatch_convex}, the result obtained by the \psctl algorithm is an exact solution for SCT-CP(G) \cite{Danisch17,FW13}. 
	\begin{corollary}
		Let $r'$ be the result obtained upon \psctl  converges. Let $\alpha'$ be the vector of weight assignment of $r'$. $\alpha'$ is the optimal solution of \ecp.
	\end{corollary}	

	\stitle{Convergence analysis of \psctl.} To analyze the convergence rate of \psctl, we use the curvature constant $C_f$~\cite{FW13} as a measure of "non-linearity" of the objective function $f(\alpha)$. The curvature constant $C_f$  of a convex and differentiable function with respect to a compact domain $\mathcal D$ is  
	\begin{equation}\label{equ:cf}\small
		C_f := \sup_{\alpha_1,\alpha_3 \in \mathcal{D}, 
			\atop {
				\gamma \in [0,1],
				\atop {
					\alpha_2=\alpha_1+\gamma(\alpha_3-\alpha_1)}}} {\frac{2}{\gamma^2}}(f(\alpha_2)-f(\alpha_1)-\langle \alpha_2-\alpha_1,\nabla f(\alpha_1) \rangle ).
	\end{equation}

	\begin{lemma}
		$C_f\le x_{max}|\C_k|^2$ where $x_{max}=\max_{u\in V} |\{P \in \P| u \in V(P)\}|$, the maximum number of pairs covering the same node in $G$.
	\end{lemma}
	\begin{proof}
		Let $Diam(\mathcal{D})$ be the squared Euclidean diameter of $\mathcal{D}$, i.e. $Diam(\mathcal{D}) := \sup_{\alpha_1,\alpha_2\in \mathcal{D}}{d(\alpha_1, \alpha_2)}$. $L= \sup_{\alpha \in \mathcal{D}}{\| \nabla^2 f(\alpha) \|_2}$ is the Lipschitz constant of $f(\alpha)$.  $\| \nabla^2 f(\alpha) \|_2$ is the spectral norm of the Hessian matrix of $f(\alpha)$ \cite{Danisch17, FW13}. 
		Here $L$ is bounded by $x_{max}$. Because $\alpha \geq 0$ and $||\alpha||_1 = |\C_k(V)|$, we have $\max_{\alpha \in \mathcal{D}}{\langle \alpha, \alpha \rangle} \le |\C_k(V)|^2$ and $Diam(\mathcal{D})^2 $ is bounded by $\max_{\alpha \in \mathcal{D}}{\langle \alpha, \alpha \rangle} \le |\C_k(V)|^2$. According to \cite{FW13}, $C_f$ is bounded by $Diam(\mathcal{D})^2 L$, thus we have $C_f\le x_{max}|\C_k|^2 $. 
	\end{proof}

	\begin{theorem}
		Let $s^t=\alpha^t - \alpha^{t-1}$ where $\alpha^t$ is the results of \psctl in $t$-th iterations. Let $s^* = \alpha^* - \alpha^{t-1}$ be difference between $\alpha^{t-1}$ and the optimal solution $\alpha^*$. We have 
		$\langle\ s^t, \nabla f(\alpha^t) \rangle - \langle\ s^*, \nabla f(\alpha^t) \rangle \le \frac{1}{2}\beta \gamma_t C_f$ where $\gamma_t=\frac{1}{t}, \beta  = \frac{4\sqrt{k} \Delta }{\sqrt{|\C_k|}}$. $\Delta$ is the maximum number of $k$-cliques that can cover one node $u\in V$. 
	\end{theorem}
	\begin{proof}
		Observe that $\frac{\alpha^t}{t}=\frac{t-1}{t} \frac{\alpha^{t-1}}{t-1} + \frac{s^t}{t}$. The update can be seen as $x^{(t)}=(1-\gamma_t) x^{(t-1)} + \gamma_t s^t $ where $x^{(t)}$ denotes$ \frac{\alpha^t}{t}$. 
		Dimension of $\alpha$ is $\sum_{P(V_h, V_p)\in \P} |V(P)|\le k|\C_k(V)|$, and we have $\| s^t-s^*\| \le \|\alpha^t\| + \|\alpha^*\| \leq {2|\C_k|}$. We have $\frac{\partial f(\alpha)}{\partial \alpha^P_u} = 2r(u)$ and $r(u)$ will be increased by at most $\Delta$ (\cond{4}, \ecp(G), the bound is given by the number of cliques in $P$ covering $u \in V_p$ and \cond{3}, \ecp(G) the bound is given by the number of cliques in $P$ covering $u\in V_h$).

	Denote by ${s}^{t,i}$ the vector of $i_{th}$ pair. Denote by $\alpha^{t,i} := \alpha^t + \sum_{j=1}^{i-1} {s^{t, i}}$. For $t\ge 1$, we have 
			\begin{equation}\tiny
				\begin{aligned}
					& \langle\ s^t, \nabla f(\frac{\alpha^t}{t}) \rangle - \langle\ s^*, \nabla f(\frac{\alpha^t}{t}) \rangle 
					\\ & = \frac{1}{t}\langle\ s^t-s^*, \nabla f(\alpha^t) \rangle 
					 = \frac{1}{t} \sum_{i=1}^{\eta} {\langle\ {s}^{t,i}-{s}^{*,i}, \nabla_{P_i} f({\alpha}^{t}) \rangle }
					\\ & = \frac{1}{t} \sum_{i=1}^{\eta} {\langle\ {s}^{t,i}-{s}^{*,i}, \nabla_{P_i} f(\alpha^t-{\alpha}^{t,i}) \rangle }
					 + \frac{1}{t} \sum_{i=1}^{\eta} {\langle\ {s}^{t,i}-{s}^{*,i}, \nabla_{P_i} f({\alpha}^{t,i}) \rangle }
					\\ & \le \frac{1}{t} \sum_{i=1}^{\eta} {\langle\ {s}^{t,i}-{s}^{*,i}, \nabla_{P_i} f(\alpha^t-{\alpha}^{t,i}) \rangle }
					\\ & = \frac{1}{t} \sum_{i=1}^{\eta} {\langle\ {s}^{t,i}-{s}^{*,i}, \nabla_{P_i} f(\alpha^t) - \nabla_{P_i} f({\alpha}^{t,i}) \rangle }
					\\ & =  \frac{1}{t}{\langle\ {s}^{t}-{s}^{*}, \nabla f(\alpha^t) - \left( \nabla_{P_1} f({\alpha}^{t,1}),...,\nabla_{P_{\eta}} f({\alpha}^{t,\eta}) \right)  \rangle }
					\\ & \le \frac{1}{t} \|  {s}^{t}-{s}^{*} \| \cdot 
					  \|\nabla f(\alpha^t) - \left( \nabla_{P_1} f({\alpha}^{t,1}), ...,\nabla_{P_{\eta}} f({\alpha}^{t,\eta}) \right) \|
					\\ & \le \frac{1}{t}\cdot \sqrt{k|\C_k(V)|} \cdot 2\Delta\cdot 2|\C_k(V)|.
				\end{aligned}
			\end{equation}
	The above  derivation is correct because \psctl can consider each pair independently. Thus, according to the Algorithm~2 in \cite{FW13}, $$\beta = \frac{1}{t} \sqrt{k|\C_k|} 2\Delta 2|\C_k| / \left(\frac{1}{2} \gamma_t  C_f \right)
		= \frac{8\sqrt{k}|\C_k(V)|^{1.5} \Delta}{  C_f} \le  \frac{4\sqrt{k} \Delta }{\sqrt{|\C_k(V)|}}.$$
		
		The last inequality comes directly from the definition of $C_f$ in Equation~\ref{equ:cf} from which we have $C_f \ge 2|\C_k|^2$. 
	\end{proof}

	\begin{theorem}
		For each $t \ge 1$, 
		$f(\alpha^t) - f(\alpha^*) \le \frac{2x_{max}|\C_k|^2}{t+2}(1 + \beta)$ where $\beta=\frac{4\sqrt{k} \Delta }{\sqrt{|\C_k(V)|}}$.
	\end{theorem}
	\begin{proof}
		The proof can be obtained based on the results established in \cite{FW13}. Specifically, in \cite{FW13}, the authors show that for each $t \ge 1$, 
		$f(\alpha^t) - f(\alpha^*) \le \frac{2C_f}{t+2}(1 + \beta)$. As describe above, $C_f$ is bounded by $x_{max}|\C_k|^2$ and $\beta$ is bounded by $\frac{4\sqrt{k} \Delta }{\sqrt{|\C_k(V)|}}$.
	\end{proof}

\section{New Sampling-Based Algorithm}
	
	Since the hardness of \kcdsp, FW-based solutions may still be costly when handling very large graphs. To further improve the efficiency, sampling-based solutions are often used which can typically obtain a good approximation of the $k$-clique densest subgraph \cite{MitzenmacherPPT15, kclpp, sctl}.
	In this section, we propose a new but efficient sampling-based  algorithm, called  \spath, which employs the \ccpath algorithm porposed in  \cite{ccpath} to sample $k$-cliques. A remarkable feature of \spath is that it has polynomial time complexity. To the best of our knowledge,  \spath is the first algorithm that  runs in polynomial time, thus it can handle large graphs. We  will also present a detailed theoretical analysis of the accuracy bound of \spath.
	

	\subsection{The \spath algorithm}
	For sampling-bases solutions, the state-of-the-arts are the \skcl and \ssctl algorithm \cite{kclpp, sctl}. \skcl counts all $k$-cliques at first, and then samples $k$-cliques uniformly. \ssctl builds the SCT-Index at first, and then samples $k$-cliques uniformly from the SCT-Index. Thus, both \skcl and \ssctl suffer from exponential time complexity and are often intractable for handling large graphs. To overcome these limitations, we develop a poly-nominal algorithm, called \spath. 
	
	First, we briefly introduce the \ccpath algorithm which was originally proposed to estimate the number of $k$-cliques in a graph \cite{ccpath}. In \spath, we use \ccpath as a uniformly $k$-clique sampler with polynomial running time. \ccpath first colors the graph (using a linear-time greedy graph coloring algorithm) such that the vertices of each edge in the graph must have different colors. \ccpath is an efficient algorithm for counting $k$-cliques through sampling from a combinatoric  structure called $k$-color path.  Specifically, a $k$-color path is a path with $k$ vertices, and the $k$ vertices have $k$ different colors. Each $k$-clique must be a $k$-color path, i.e. the set of $k$-cliques is a subset of $k$-color paths  (because the vertices in a clique must have different colors). \ccpath is a polynomial dynamic programming algorithm which can uniformly samples from  $k$-color paths. Since the set of $k$-cliques is a subset of $k$-color paths, the $k$-cliques can also be  sampled uniformly. Note that checking if a $k$-color-path is a $k$-clique can be easily done in $O(k^2)$ time by verifying whether any pair of node is connected. Similar to \cite{ccpath}, we can regard the probability $p'$ that a CCPATH is a $k$-clique as a graph-related parameter which is often very  high as shown in \cite{ccpath}.   
	
	The details of \spath are outlined in Algorithm~\ref{alg:sample}. Algorithm~\ref{alg:sample} admits $t$ as a parameter for the size of samples (line~1). Firstly, the algorithm  uniformly samples  the $k$-cliques and get the set of samples $\mathbb{C}$ (lines~1-2). Then, the algorithm invokes  \kcl on $\mathbb{C}$ in a small number of iterations, and  then returns the approximation result (lines~3-10). 
	
	\begin{algorithm}[t]
		\caption{The Proposed \spath Algorithm}
		\label{alg:sample}
		\small
		\KwIn{The graph $G(V, E)$, clique size $k$, sample size $t$, a smaller number $T$}
		\KwOut{An approximation of $V^*$}
		\SetKwProg{Fn}{Procedure}{}{}
		
		$\mathbb{P}\gets$ get $t$ uniform samples from all  $k$-color paths through the \ccpath algorithm\;
		$\mathbb{C}\gets $ the set of $k$-cliques in $\mathbb{P}$\; 
		
		$r(u)\gets 0, \forall u\in V$\;
		\For{$t\gets 1$ to $T$}{
			\ForEach{$k$-clique $C\in \mathbb{C}$}{
				$u\gets \arg \min_{v\in C}{r(v)}$\;
				$r(u)\gets r(u)+1$\;
			}
		}	
		Sort $V$ by $r(u)$ \textcolor{black}{in non-increasing order}\;
		Denote $V_i$ by the previous $i$ vertices of \textcolor{black}{the sorted} $V$\;
		\Return{$\arg\max_{V_i} \rho_k'(V_i)$ where $\rho_k'(V_i)=\frac{|\C_k(V_i)\cap \mathbb{C}|}{|V_i|}$}\;
	\end{algorithm}
	
	As shown in Algorithm~\ref{alg:sample}, the proposed \spath  algorithm is very simple, but it is very efficient.  Below, we first analyze the time and space complexity of Algorithm~\ref{alg:sample}.
	 
	
	\begin{theorem}\label{the:sample_complexity}
		The time complexity of Algorithm~\ref{alg:sample} is $O(|V|\delta^2k+(\delta k + k^2)t + Tkt)$. And the space complexity of Algorithm~\ref{alg:sample} is $O(kt)$.
	\end{theorem}
	\begin{proof}
		Note that by \cite{ccpath} the time complexity taken by sampling is $O(|V|\delta^2k+(\delta k + k^2)t)$ \cite{ccpath}. Since the size of $\mathbb{C}$ is bounded by $t$, the time complexity of \kcl is $O(Tkt)$. The memory cost of Algorithm~\ref{alg:sample}  is dominated by storing of $\mathbb{C}$, which uses at most $O(kt)$ space.
	\end{proof}
	
	As shown in Theorem~\ref{the:sample_complexity}, \spath is completely free from the count of $k$-cliques. As shown in our experiments, \spath is very efficient and it can solve the $k$-DSS problem on very large graphs that has vertices more than 1.8 $\times 10^9$. Below, we present the theoretical analysis of the accuracy guarantee of the \spath algorithm. \textcolor{black}{The results show that the accuracy of \spath is  based on a mild condition, which also explains the good accuracy  of the previous works \cite{kclpp, sctl}. }

	\subsection{Analysis of the Algorithm}
	In this subsection, we analyze the accuracy bound of \spath. For our analysis, we need the following Chernoff bound.
	
	\begin{theorem}[Chernoff bound]
		Let $X=\sum_{i=1}^{n}{X_i}$, where $X_i$ are independent with each other. $X_i=1$ with probability $p_i$ and $X_i=0$ with probability $1-p_i$. Let $E[X]=\mu=\sum_{i=1}^{n}{p_i}$. $\theta$ is a number that $0<\theta < 1$. Then 
		$$\Pr(|X-\mu|\ge \theta \mu ) \le 2\exp(-\frac{\mu \theta^2}{3}).$$
	\end{theorem}
	
	Let $R$ be the set of vertices $\cup_{C\in \mathbb{C}}{C}$, where $\mathbb{C}$ is the set of sampled $k$-cliques (line~2 of Algorithm~\ref{alg:sample}). Let $G(R^*)$ be the ground truth of the $k$-clique densest subgraph in $G(R)$, which has the largest value $\rho_k$. Let $\C'_k(U)=|\C_k(U)\cap \mathbb{C}|$ and  $\rho_k'(U)=\frac{|\C'_k(U)|}{|U|}$ denotes the $k$-clique density in the sampled $k$-cliques. $\tilde{R^*}$ is the result returned by \spath, which is expected to have the largest value of $\rho_k'$.
	
	Let $s(C)$ be the random variable that indicates whether the $k$-clique $C$ is sampled or not, i.e. $s(C)=1$ if $C\in \mathbb{C}$ or $s(C)=0$ otherwise. Let $p$ be the probability of each $k$-clique being sampled. {Denote $p'$ by the probability of a $k$-color path being a $k$-clique (line~2 of Algorithm~\ref{alg:sample}), then we have $p=\frac{tp'}{|\C_k(V)|}$.}  Subsequently, we can derive the following results. 
	

	\begin{lemma}\small
		$E[\C'_k(U)]=p\C_k(U)$.
	\end{lemma}
	\begin{proof}\small
		$E[\C'_k(U)]=E[|\C_k(U)\cap\mathbb{C}|]=E[\sum_{C\in \C_k(U)}{s(C)}]=p\C_k(U)$.
	\end{proof}
	
	\begin{lemma}\label{lem:expec}\small
		$E[\rho_k'(U)]=p\rho_k(U)$.
	\end{lemma}
	\begin{proof}\small
		
		$E[\rho_k'(U)]=E[\frac{|\C'_k(U)|}{|U|}]=\frac{p\C_k(U)}{|U|}=p\rho_k(U)$.
		
	\end{proof}

Based on the above lemmas, we can obtain the following theorem. 
	
	\begin{theorem}\label{the:accuracy}
		Let $\epsilon$ and $\theta$ be small constant numbers.  We can conclude that $\rho_k(\tilde{R^*})$ is an $1-2\theta$ approximation of $\rho_k(V^*)$ with probability $1-\epsilon$ if \textcolor{black}{$\rho_k(\tilde{R^*})\ge \frac{-3\ln{\epsilon/4}}{p\theta^2}$.}  
	\end{theorem}
	
	\begin{proof}
		Set the vertices set $U$ in Lemma~\ref{lem:expec} by $V^*$. Then, we have $E[\rho_k'(V^*)]=p\rho_k(V^*)$. 
		According to Chernoff bound, we have 
		\begin{equation}\small
		\Pr(|\rho_k'(V^*) - p\rho_k(V^*)|\ge \theta  p\rho_k(V^*) ) \le 2\exp(-\frac{p\rho_k(V^*) \theta^2}{3}).
		\nonumber
		\end{equation}

		This immediately gives 
		$$\Pr(p\rho_k(V^*)-\rho_k'(V^*) \ge \theta  p\rho_k(V^*) ) \le 2\exp(-\frac{p\rho_k(V^*) \theta^2}{3}).$$
		Since $\rho_k'(\tilde{R^*})\ge\rho_k'(V^*)$, then  we have
		\begin{equation}\label{equ:1}
			\Pr(p\rho_k(V^*)-\rho_k'(\tilde{R^*}) \ge \theta  p\rho_k(V^*) ) \le 2\exp(-\frac{p\rho_k(V^*) \theta^2}{3}).
		\end{equation}
		
		Let the vertices set $U$ in Lemma~\ref{lem:expec} be $\tilde{R^*}$, we have 
		$$\Pr(|\rho_k'(\tilde{R^*}) - p\rho_k(\tilde{R^*})|\ge \theta  p\rho_k(\tilde{R^*}) ) \le 2\exp(-\frac{p\rho_k(\tilde{R^*}) \theta^2}{3}).$$
		As a result, we have
		
		$$	\Pr(\rho_k'(\tilde{R^*}) - p\rho_k(\tilde{R^*})\ge \theta  p\rho_k(\tilde{R^*}) ) \le 2\exp(-\frac{p\rho_k(\tilde{R^*}) \theta^2}{3}).$$
		
		Since $\rho_k(V^*)\ge \rho_k(\tilde{R^*})$,  we can derive that
		\begin{equation}\label{equ:2}
			\Pr(\rho_k'(\tilde{R^*}) - p\rho_k(\tilde{R^*})\ge \theta  p\rho_k(V^*) ) \le 2\exp(-\frac{p\rho_k(\tilde{R^*}) \theta^2}{3}).
		\end{equation}
		
		With Equations~\ref{equ:1} and \ref{equ:2}, we are able to derive that 
		\begin{equation}
			\begin{aligned}
				\Pr & \left( \rho_k(V^*) - \rho_k(\tilde{R^*})  \ge 2 \theta  \rho_k(V^*)
				\right)  \le \\  & 2\exp(-\frac{p\rho_k(V^*)\theta^2}{3})+2\exp(-\frac{p\rho_k(\tilde{R^*})\theta^2}{3}).
			\end{aligned}
		\end{equation}
		
		To ensure
		$\Pr  \left( \frac{ \rho_k(V^*) - \rho_k(\tilde{R^*})}{ \rho_k(V^*)}  \ge 2 \theta 
		\right) \le \epsilon,$ it has $2\exp(-\frac{p\rho_k(V^*)\theta^2}{3})\le \frac{\epsilon}{2}$ and $2\exp(-\frac{p\rho_k(\tilde{R^*})\theta^2}{3} \le \frac{\epsilon}{2},$ and we can conclude that  $p\rho_k(V^*)\ge p\rho_k(\tilde{R^*})\ge \frac{-3\ln{\epsilon/4}}{\theta^2}$.
	\end{proof}

\begin{theorem}\label{the:expectation}
In expectation, our \spath algorithm can achieve an $(\epsilon,\theta)$-approximation if $t\ge \frac{-3|V|\ln{\epsilon/4}}{p'\theta^2}$.
\end{theorem}
\begin{proof}
		We can bound $\rho_k(\tilde{R^*})$ by $\rho_k(V)$ in Theorem~\ref{the:accuracy}, $$p\rho_k(\tilde{R^*})\ge p\rho_k(V) = \frac{tp'}{|\C_k(V)|}\frac{|\C_k(V)|}{|V|}\ge \frac{-3\ln{\epsilon/4}}{\theta^2}.$$ Thus, we have 
		$t\ge \frac{-3|V|\ln{\epsilon/4}}{p'\theta^2}.$
\end{proof}	

Theorem~\ref{the:expectation} proves that the number of iterations of \spath is liner with respect to $|V|/p'$.

\stitle{Discussions.} \textcolor{black}{ Theorem~\ref{the:accuracy} shows that if a subgraph reported by a sampling-based algorithm is dense enough, the subgraph should also be an good approximation.} Note that Theorem~\ref{the:accuracy} is based on the condition that $\rho_k(\tilde{R^*})$ should be large enough. 
It is important to note that this is a mild condition. The reasons are as follows.  First, $\frac{-3\ln{\epsilon/4}}{\theta^2}$ is not large. For example, $\frac{-3\ln{\epsilon/4}}{\theta^2}$ is around $1797$ when $\epsilon = 0.01$ and $\theta = 0.1$. Second, when $p \rho_k(V^*)$ is small, the input graph should be very sparse, thus we can utilize exact algorithms to solve it.  Third, $p$ is always not small.  As shown in \cite{ccpath}, $p'$ is often a large value for real-world graphs. \textcolor{black}{Fourth, \spath returns the subgraph with the maximum value of $\rho'_k$  (line~10 of Algorithm~\ref{alg:sample}).} Based on these reasons, such a condition is often easily satisfied for real-world graphs.

\begin{table}[t]
	\scriptsize
	\caption{Datasets ($\delta$ is the degeneracy, $\eta$ is the size of SCT)}
	\label{tab:networks}
	\centering
		\begin{tabular}{c c c c c }
			\toprule
			\textbf{Networks} & $\mathbf{|V|}$  & $\mathbf{|E|}$ & $\mathbf{\delta}$ & $\eta $  \\ 
			\midrule
			\wik & 7,115 & 100,762 & 53 & 489,803  \\
			\cai & 26,475 & 53,381 & 22 & 8,312 \\
			\epi & 75,879 & 405,740 & 67 & 1,437,313  \\
			\gow & 196,591 & 950,327 & 51 & 930,005  \\
			\ama & 403,394 & 2,443,408 & 10 & 660,944  \\
			\dbl & 425,957 & 1,049,866 & 113 & 166,725  \\
			\ber & 685,230 & 6,649,470 & 201 & 2,430,187 \\
			\you & 1,157,827 & 2,987,625 & 51 & 1,397,529  \\
			\pok & 1,632,803 & 22,301,964 & 47 & 12,492,547 \\
			\ski & 1,696,415 & 11,095,298 & 111 & 12,548,404 \\
			\ork & 3,072,627 & 117,185,083 & 253 & 264,754,163 \\
			\fri & 65,608,366 & 1,806,067,135 & 304 & 3,876,765,479 \\
			\bottomrule
		\end{tabular}
	\vspace*{-0.3cm}
\end{table}

\begin{figure*}[h]
	\begin{center}
		\begin{tabular}[t]{c}
			\subfigure[\wik]{
				\includegraphics[width=0.18\linewidth]{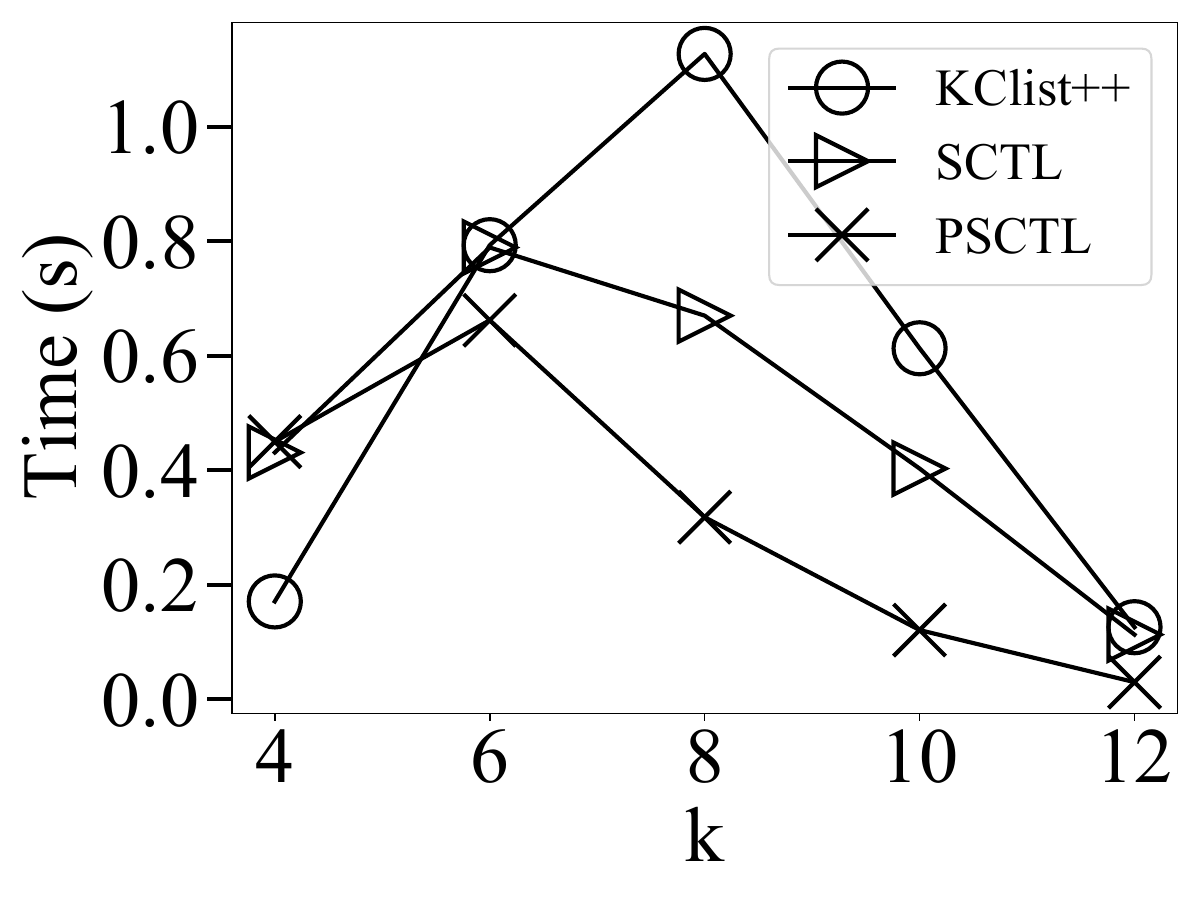}
			}
			\subfigure[\cai]{
				\includegraphics[width=0.18\linewidth]{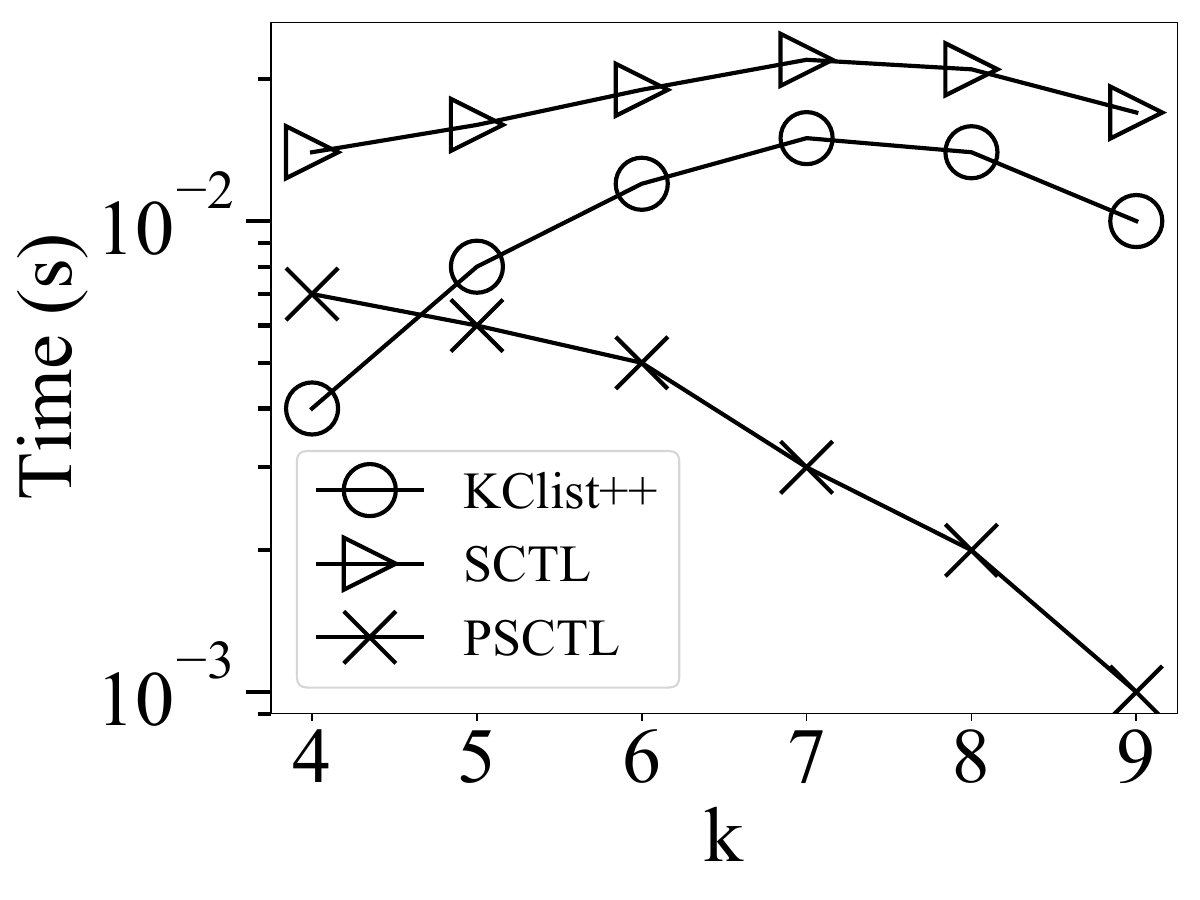}
			}
			\subfigure[\epi]{
				\includegraphics[width=0.18\linewidth]{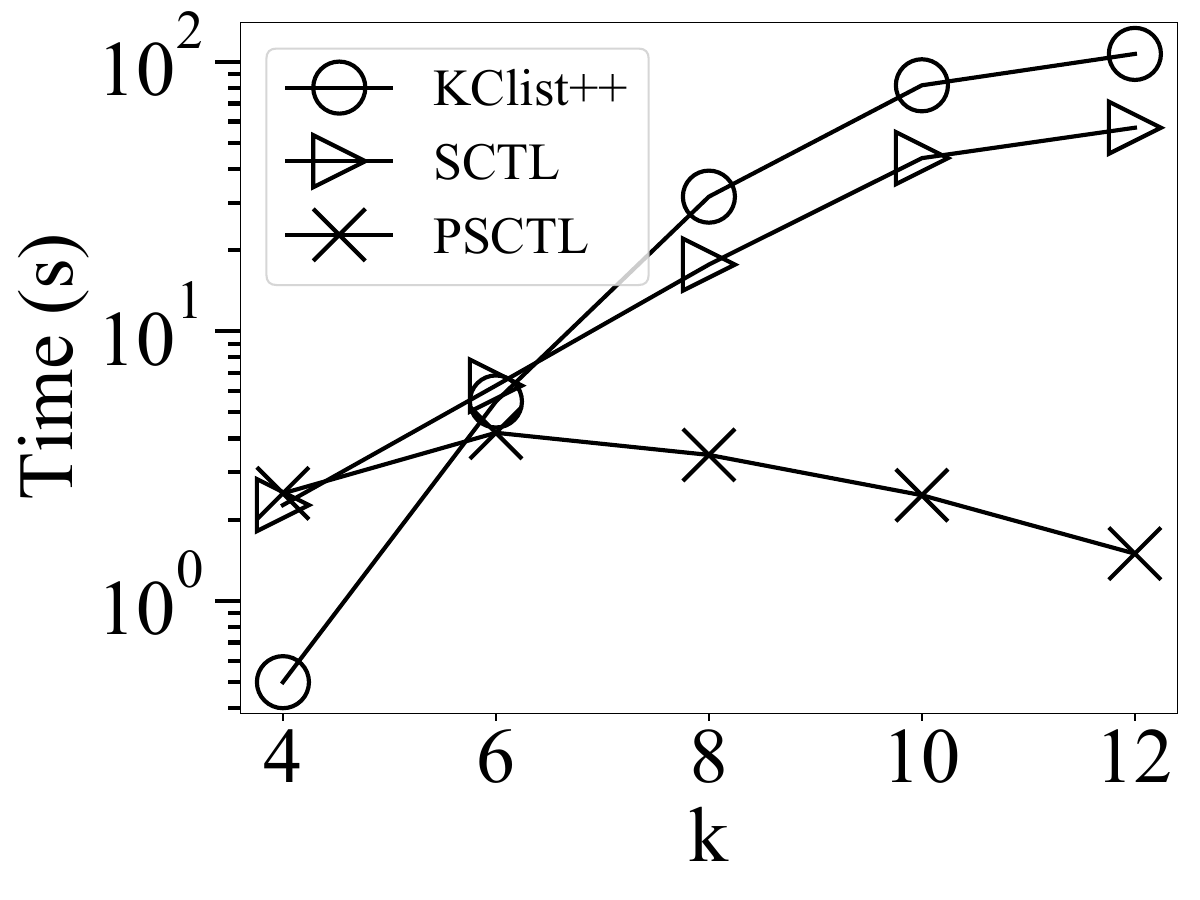}
			}
			\subfigure[\gow]{
				\includegraphics[width=0.18\linewidth]{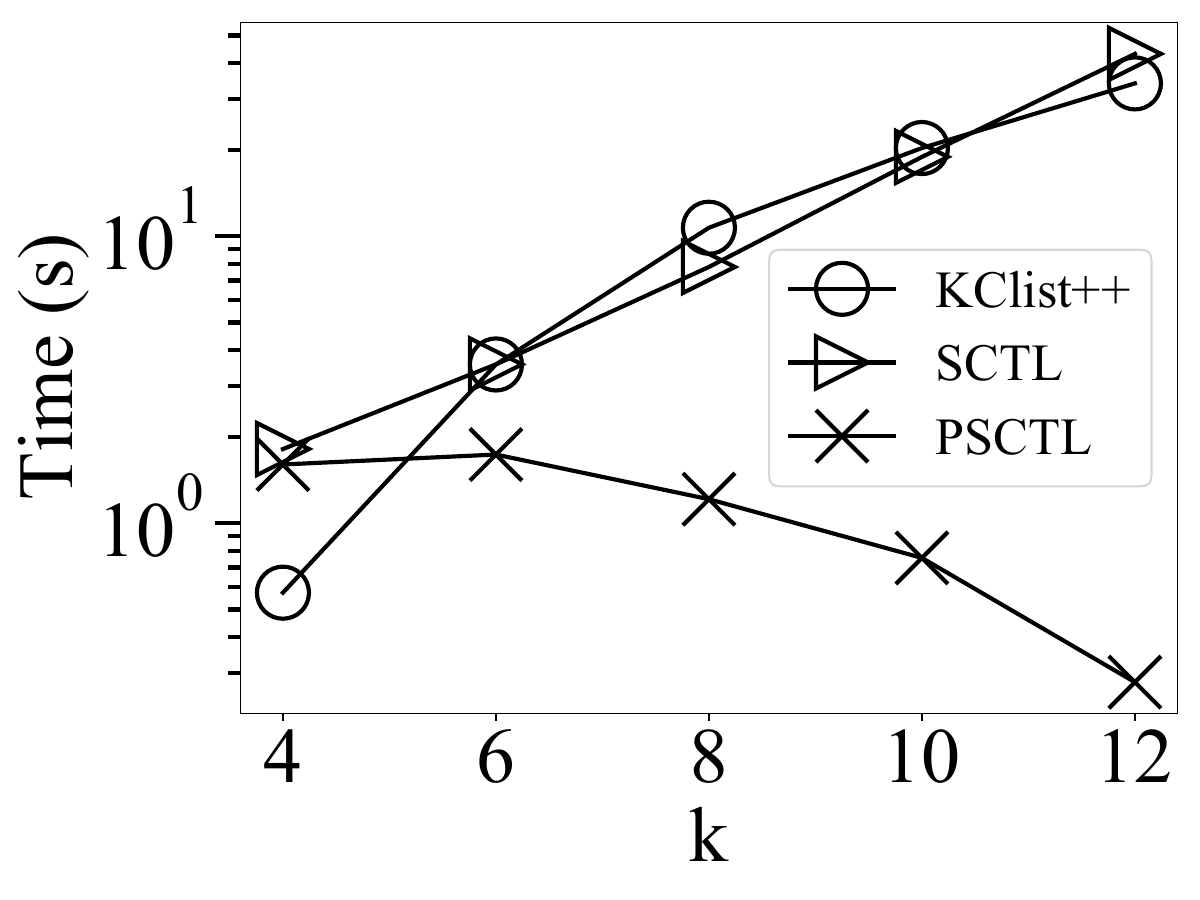}
			}
			\subfigure[\ama]{
				\includegraphics[width=0.18\linewidth]{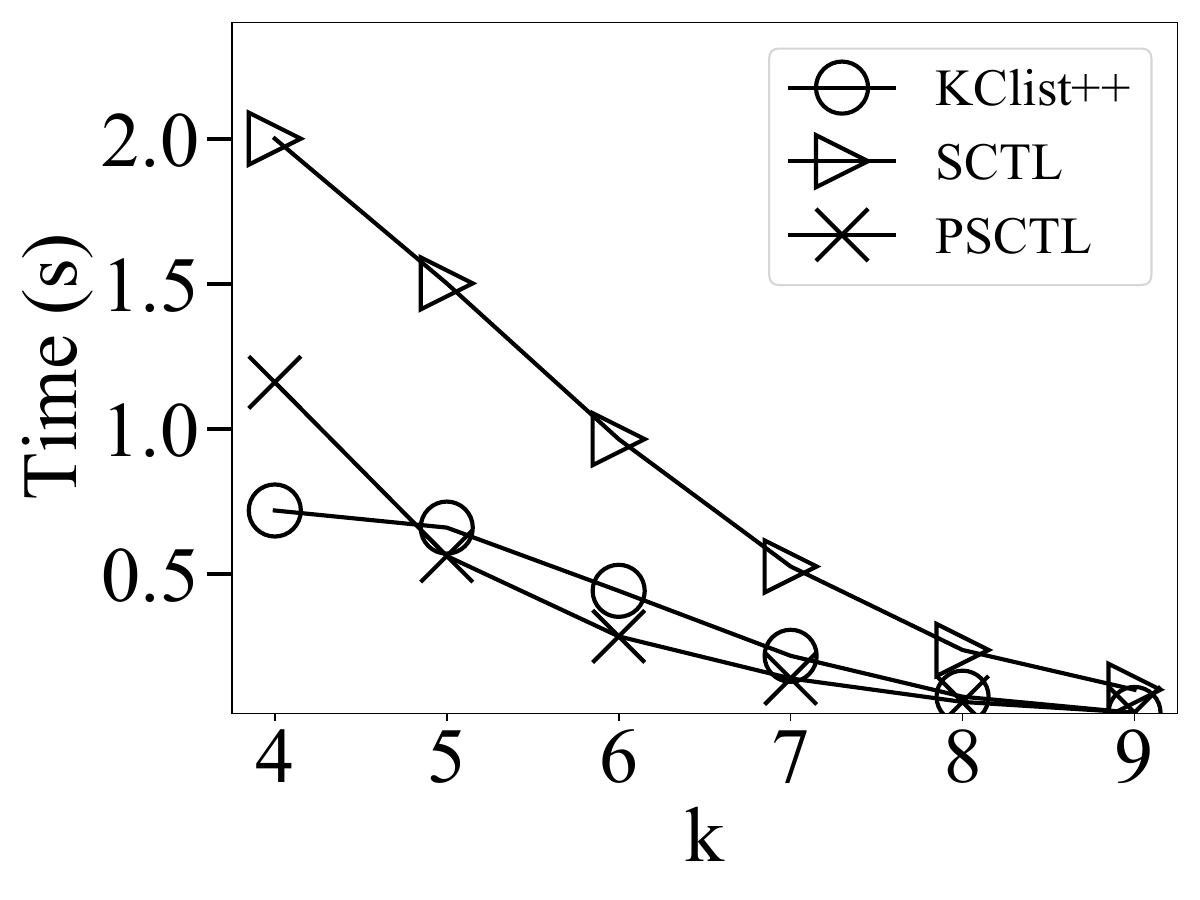}
			}
			\\
			\subfigure[\dbl]{
				\includegraphics[width=0.18\linewidth]{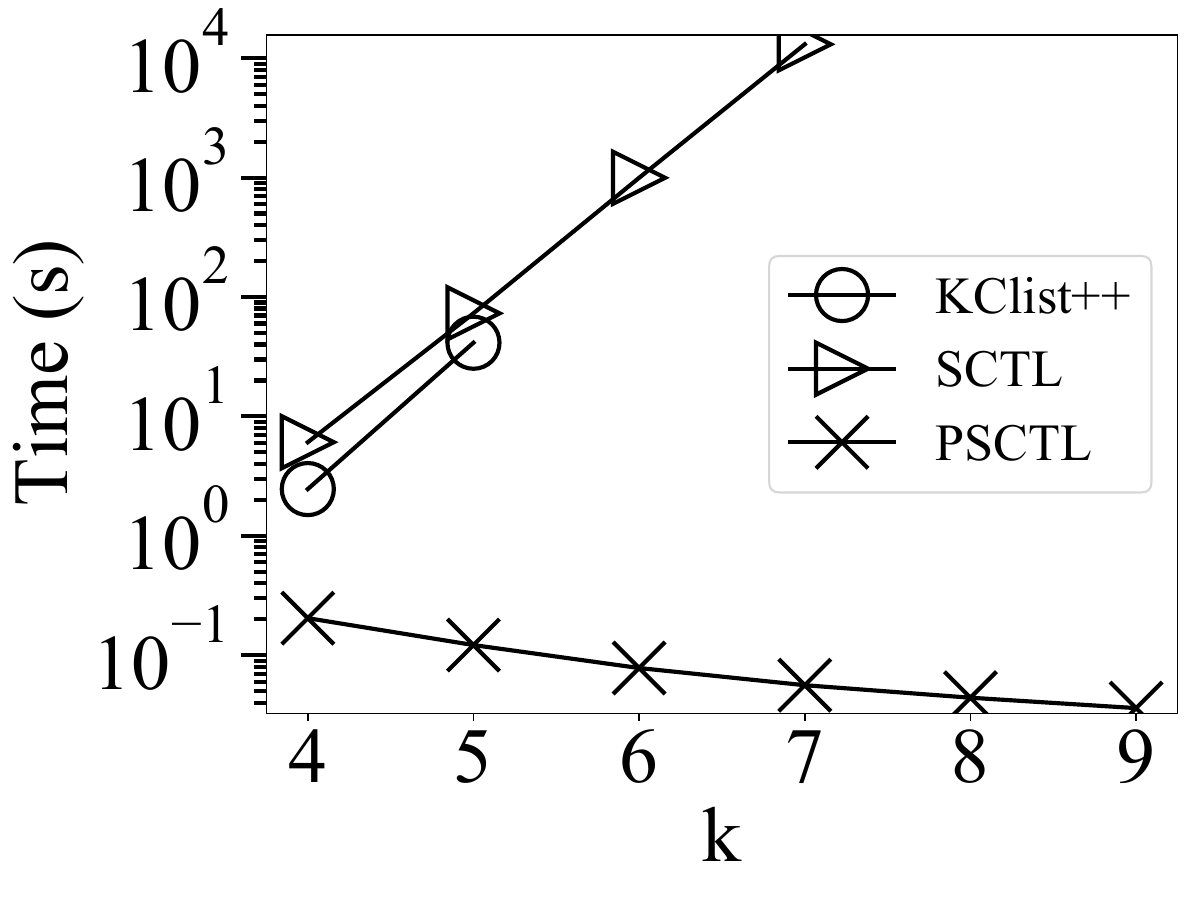}
			}
			\subfigure[\ber]{
				\includegraphics[width=0.18\linewidth]{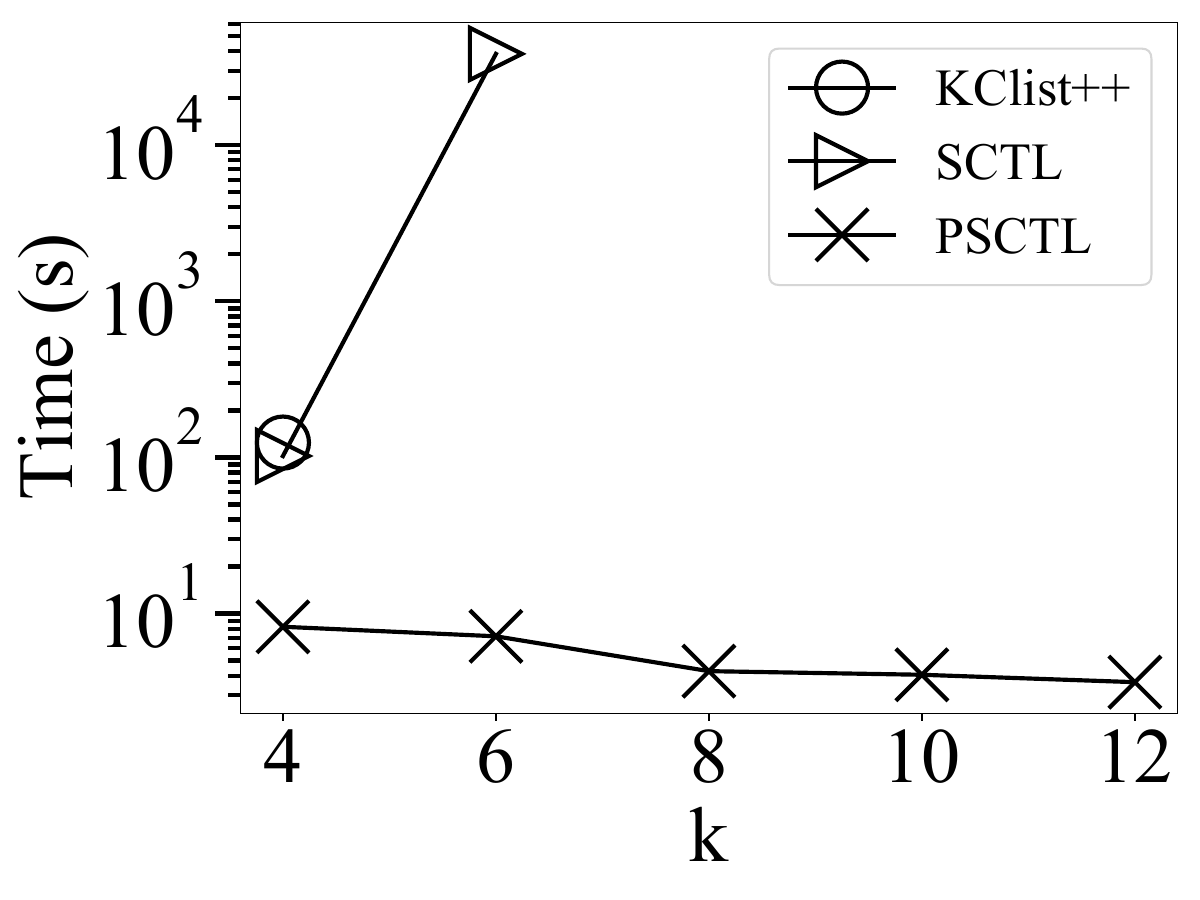}
			}
			\subfigure[\you]{
				\includegraphics[width=0.18\linewidth]{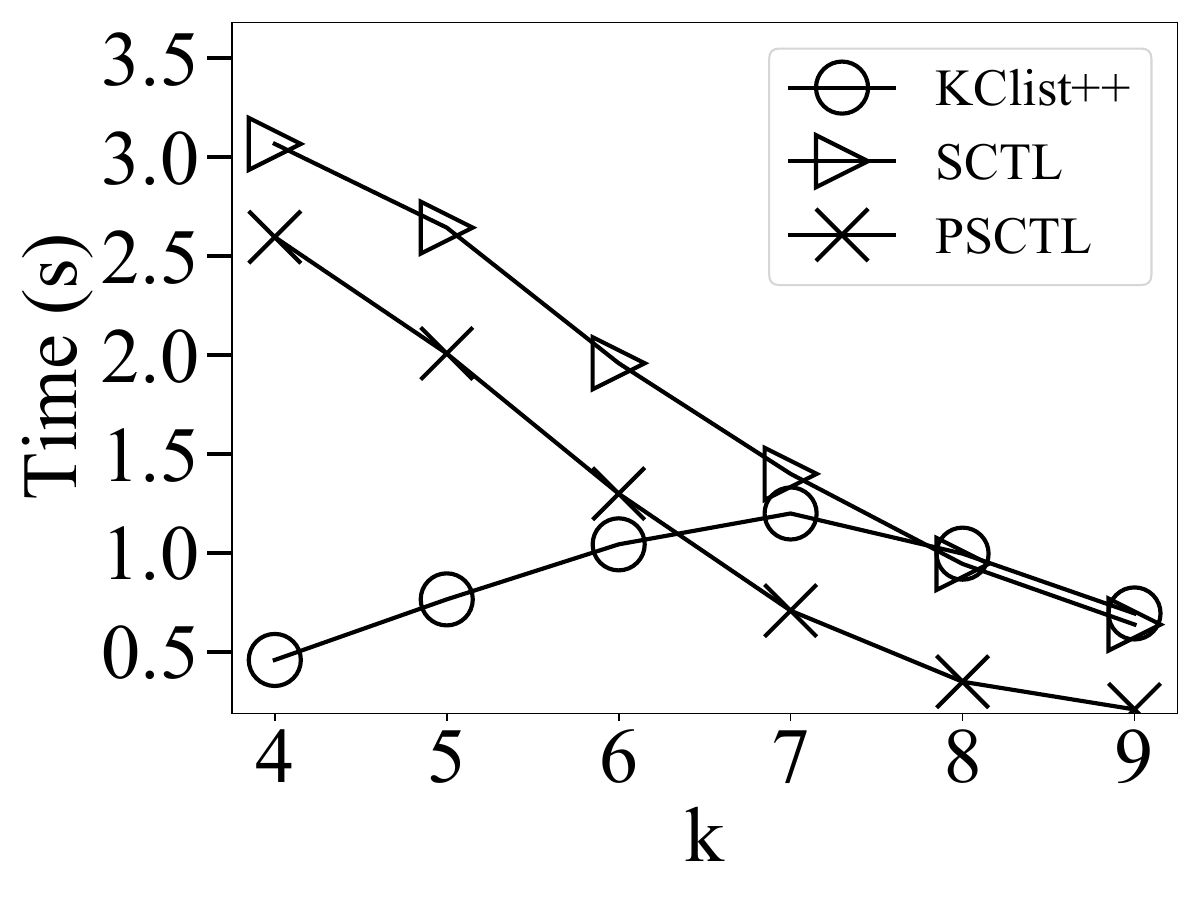}
			}
			\subfigure[\pok]{
				\includegraphics[width=0.18\linewidth]{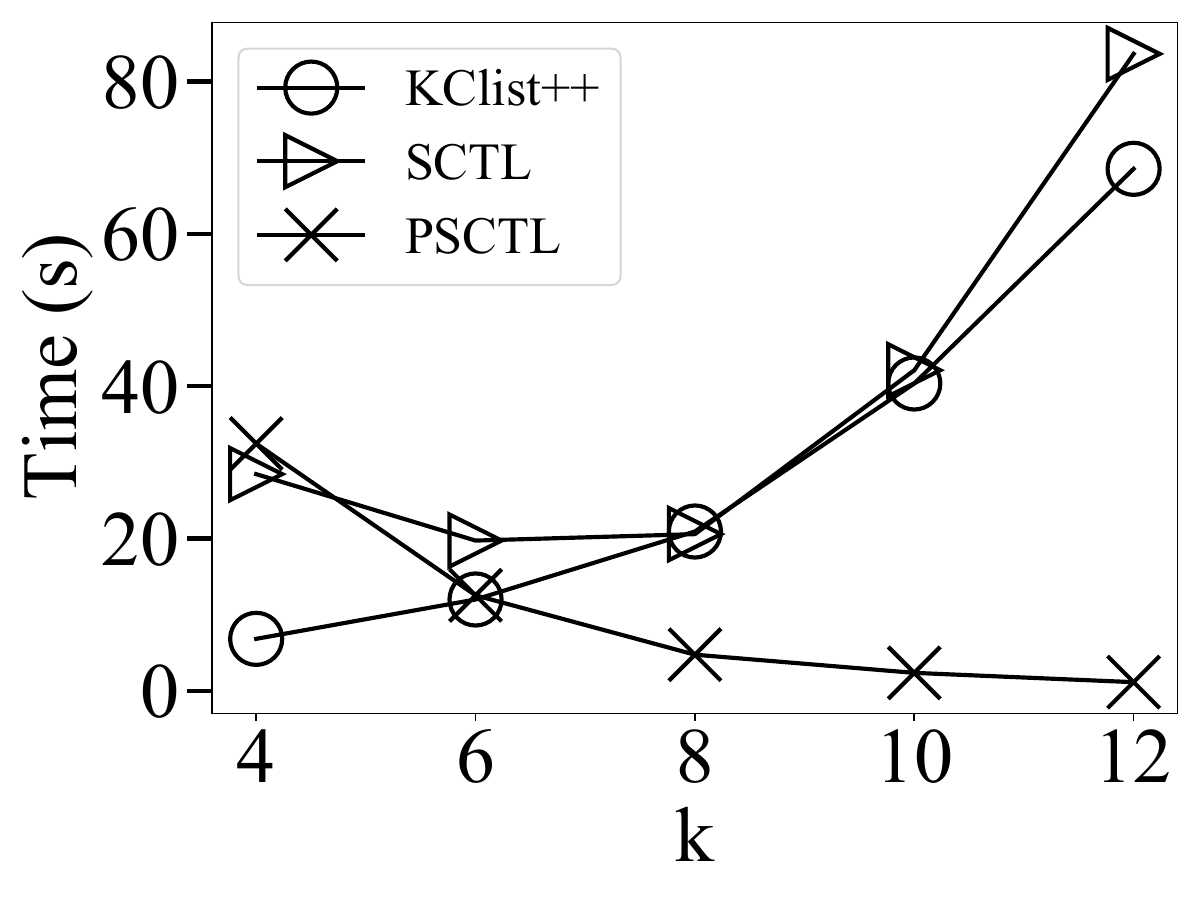}
			}
			\subfigure[\ski]{
				\includegraphics[width=0.18\linewidth]{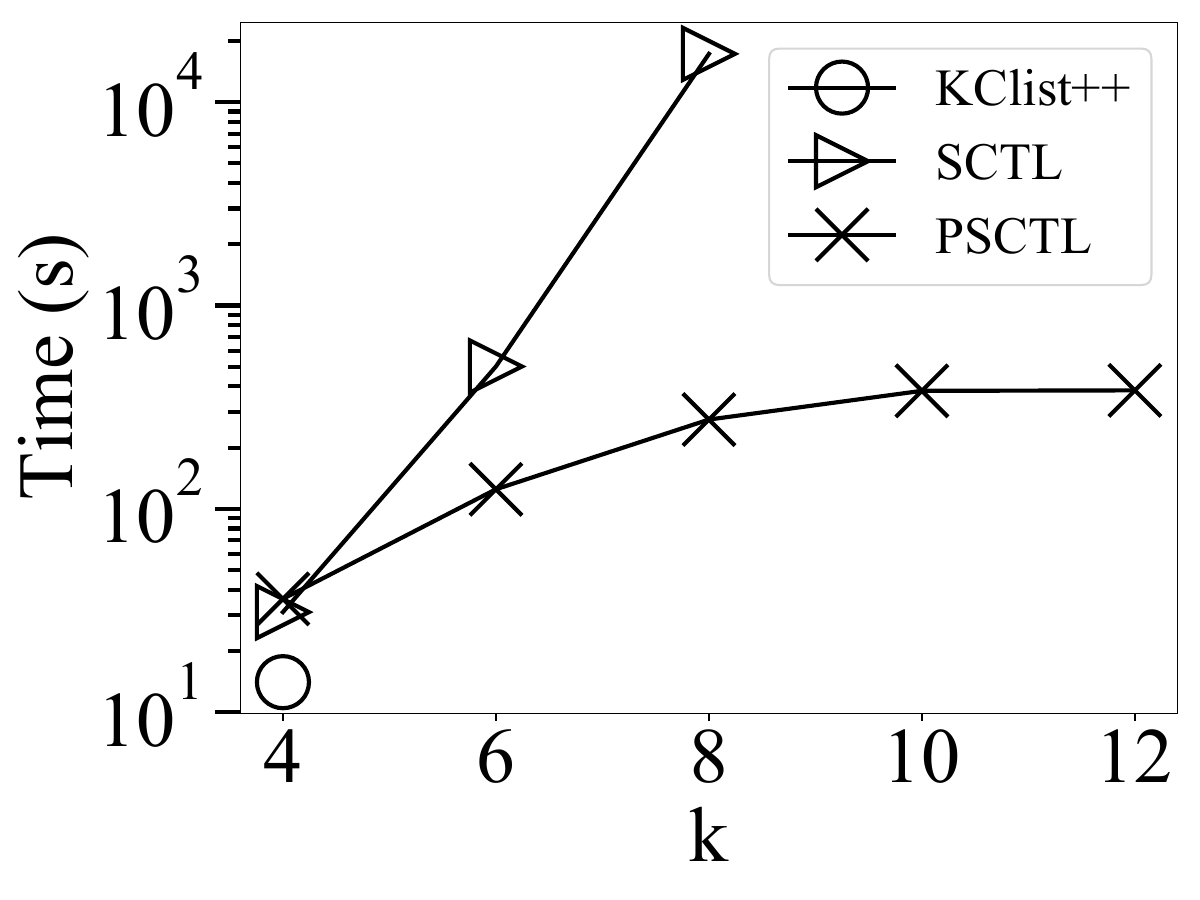}
			}
		\end{tabular}
	\end{center}
		\vspace*{-0.2cm}
	\caption{Running time of the Frank-Wolfe based algorithms ($T=10$).}
	\label{fig:fftime}
\end{figure*}

\begin{figure}
	\begin{center}
		\begin{tabular}[t]{c}
			\subfigure[\gow]{
				\includegraphics[width=0.42\linewidth]{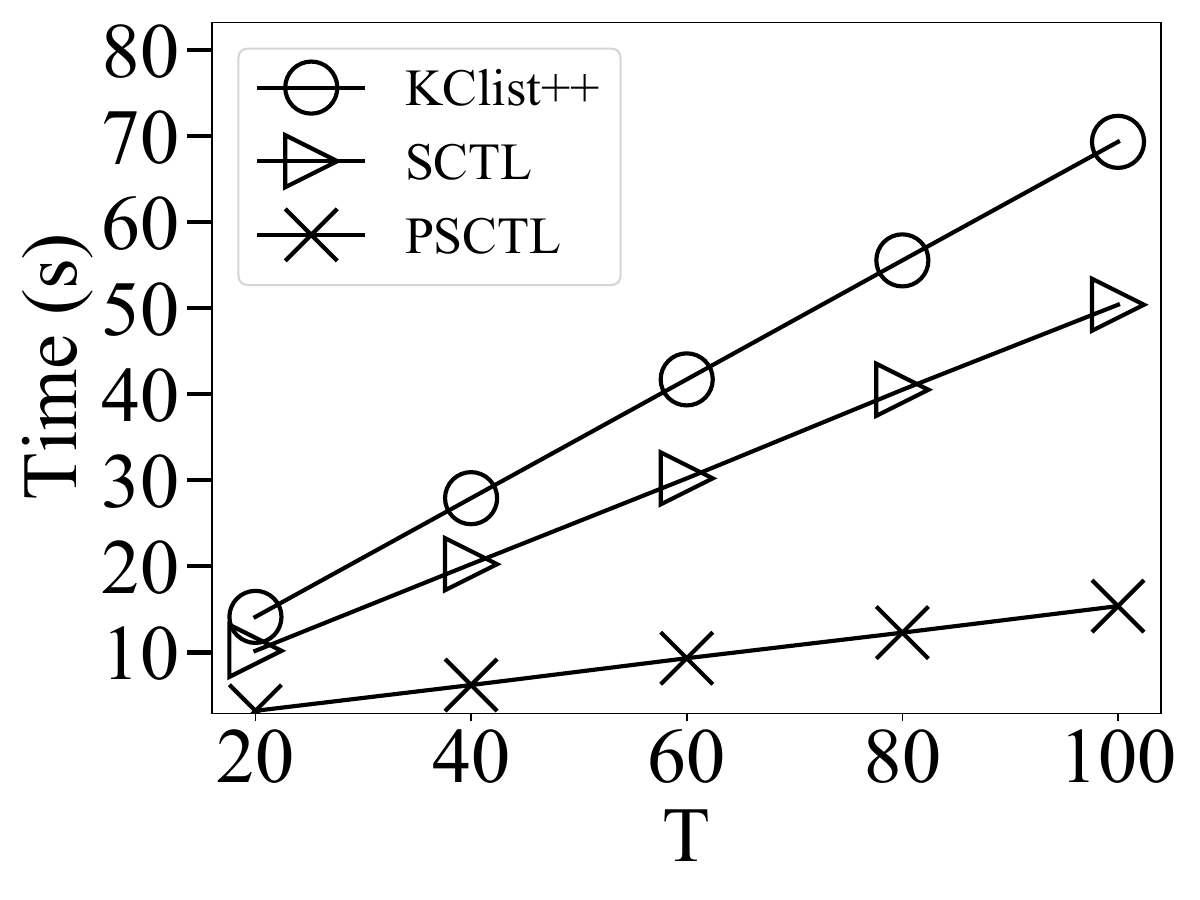}
			}
			\subfigure[\pok]{
				\includegraphics[width=0.42\linewidth]{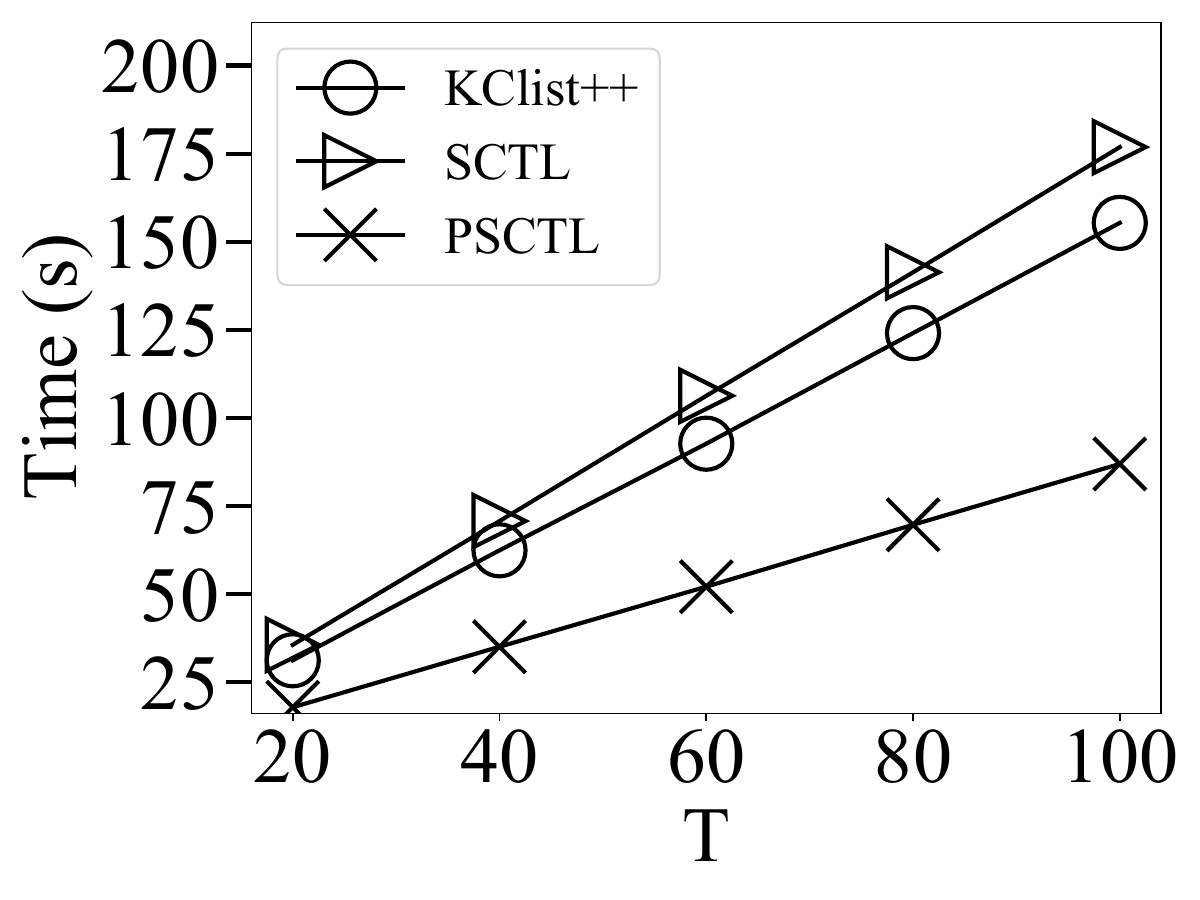}
			}
		\end{tabular}
	\end{center}
		\vspace*{-0.3cm}
	\caption{Running time of different  Frank-Wolfe based algorithms with varying $T$.}
		\vspace*{-0.3cm}
	\label{fig:vart}
\end{figure}

\begin{figure}
	\begin{center}
		\begin{tabular}[t]{c}
			\subfigure[\ama]{
				\includegraphics[width=0.42\linewidth]{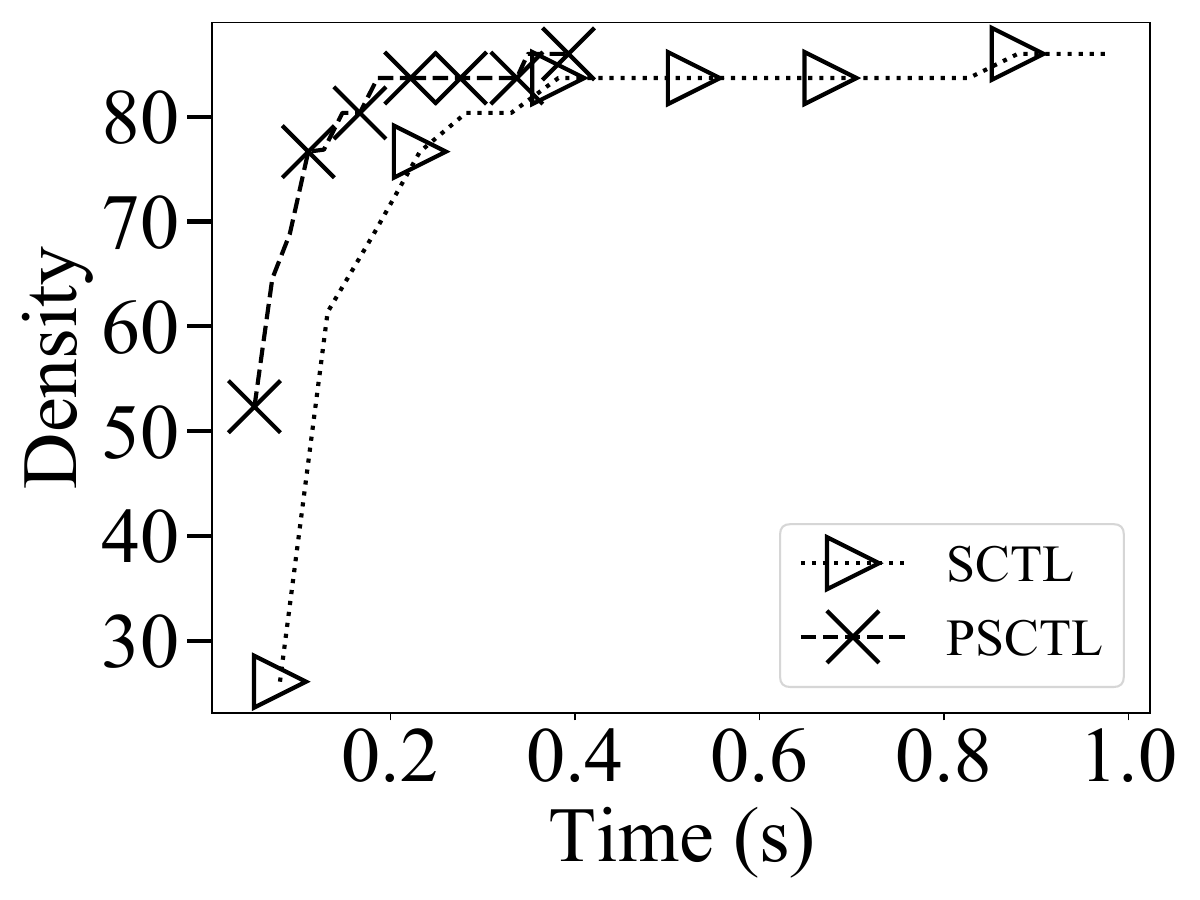}
			}
			\subfigure[\gow]{
				\includegraphics[width=0.42\linewidth]{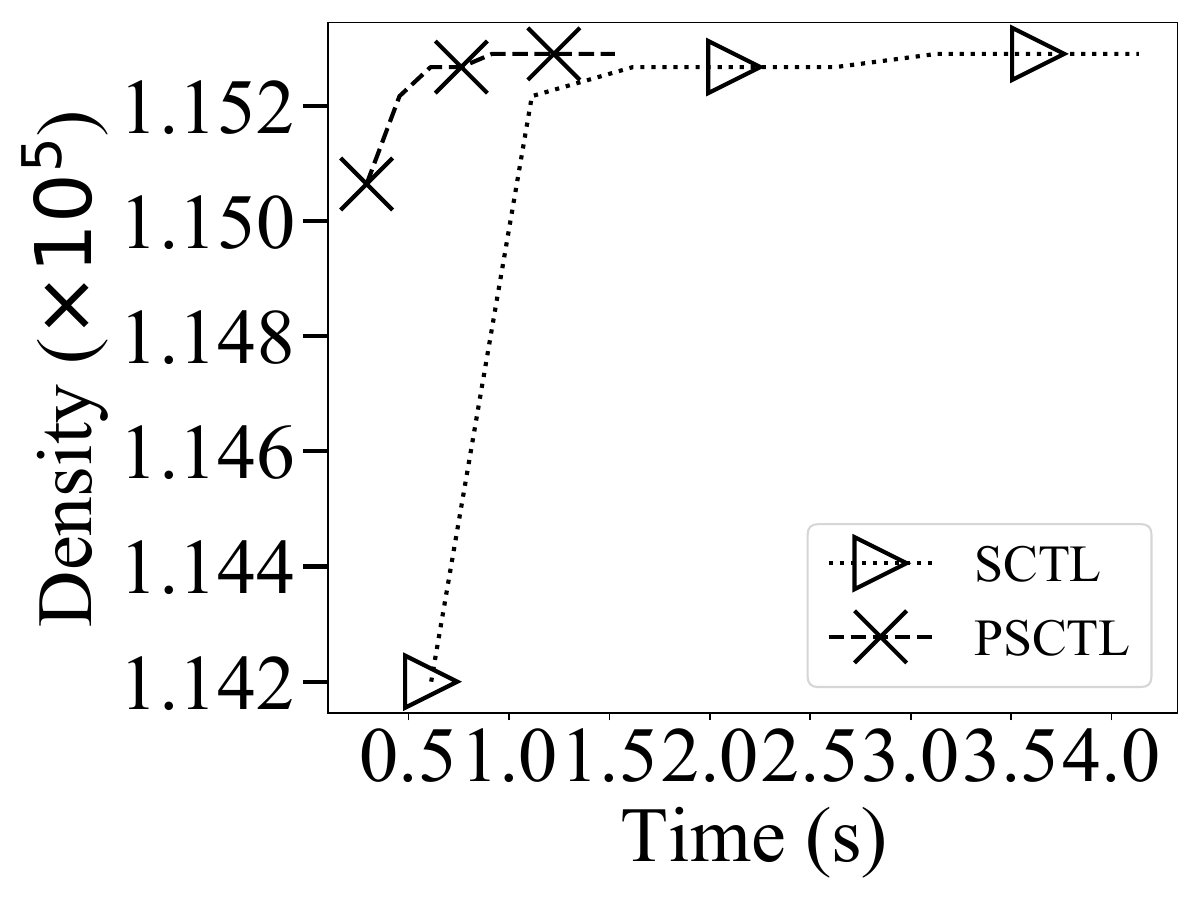}
			}
		\end{tabular}
	\end{center}
		\vspace*{-0.2cm}
	\caption{The $k$-clique density obtained by various algorithm within the same running time.}
		\vspace*{-0.3cm}
	\label{fig:fixTimeFW}
\end{figure}

%

	\begin{table*}
		\scriptsize
		\caption{The $k$-clique density for various $T$ ($k=7$).}
			 \vspace*{-0.2cm} 	
		\label{tab:convergence}
		\centering
			\begin{tabular}{c| c c c | c c c | ccc}
				\toprule
				\multirow{2}{*}{\textbf{Networks}} & \multicolumn{3}{c|}{$T=1$} & \multicolumn{3}{c|}{$T=5$} & \multicolumn{3}{c}{$T=10$}  \\
				\cmidrule{2-10}
				& \kcl  & \sctl & \psctl& \kcl  & \sctl & \psctl  & \kcl  & \sctl & \psctl \\ 
				\midrule
\wik  &  35182.25  &  35182.25  &  35182.25  &  35182.25  & 35182.25  &  35182.25  &  35182.25  &  35182.25  &  35182.25 \\
\cai  &  2203.84  &  2203.84  &  2203.84  &  2203.84  & 2203.84  &  2203.84  &  2203.84  &  2203.84  &  2203.84 \\
\epi  &  482125.39  &  481818.44  &  481818.44  &  482125.39  & 482125.39   &  482125.39  &  482125.39  &  482125.39  &  482125.39 \\
\gow  &  115291.27  &  114200.51  &  115064.38  &  115291.27  &  115268.24  &  115291.27  &  115291.27  &  115291.27  &  115291.27 \\
\ama  &  86.00  &  26.08  &  52.33  &  86.00  &  80.37  &  76.86  &  86.00  &  83.69  &  83.69 \\
\dbl  & -   &  360937368.00  &  360937368.00  & -   & 360937368.00   &  360937368.00  & -   & 360937368.00  &  360937368.00 \\
\ber  & -   & -   &  1226107478.17  & -   & -   & 1230103452.99  & -   & -   &  1230103452.99 \\
\you  &  15137.78  &  15045.44  &  15044.36  &  15137.78  &  15130.52  & 15116.24  &  15137.78  &  15134.27  &  15127.43 \\
\pok  &  137917.47  &  137917.47  &  137917.47  &  137917.47  &  137917.47  &  137917.47  &  137917.47  &  137917.47  &  137917.47 \\
\ski  & -   &  111767674.10  &  111767674.13  & -   & 111861828.44   & 111861828.44   & -   &  111882281.10  &  111882281.10 \\
				
				\bottomrule
			\end{tabular}
		\vspace*{-0.2cm}
	\end{table*}

	\begin{table}
		\scriptsize
		\caption{The number of iterations and running time needed to find $V^*$ ($k=7$).}
			 \vspace*{-0.3cm} 	
		\label{tab:convergence2}
		\centering
			\begin{tabular}{c| c |c| c| c|c|c }
				\toprule
				\multirow{2}{*}{\textbf{Networks}	} & \multicolumn{2}{c|}{\kcl } &  \multicolumn{2}{c|}{\sctl } &  \multicolumn{2}{c|}{\psctl } \\ 
				\cmidrule{2-7}
				& $T$ & Time (s) & $T$ & Time (s) & $T$ & Time (s) \\ 
				\midrule
				\wik  &  1  &  0.87  &  1  &  0.14  &  1  &  0.12 \\
				\cai  &  1  &  0.01  &  1  &  0.01  &  1  &  0.00 \\
				\epi  &  1  &  11.61  &  2  &  2.29  &  2  &  1.17 \\
				\gow  &  1  &  5.34  &  5  &  2.67  &  5  &  0.96 \\
				\ama  &  1  &  0.13  &  20  &  1.55  &  20  &  0.45 \\
				\dbl  &  -  & -   &  1  &  1286.87  &  1  &  0.04 \\
				\you  &  1  &  0.91  &  20  &  2.77  &  110  &  14.88 \\
				\pok  &  1  &  10.40  &  1  &  2.47  &  1  &  1.62 \\
				\bottomrule
			\end{tabular}
	\end{table}

	\begin{figure}
		\begin{center}
			\includegraphics[width=0.7\linewidth]{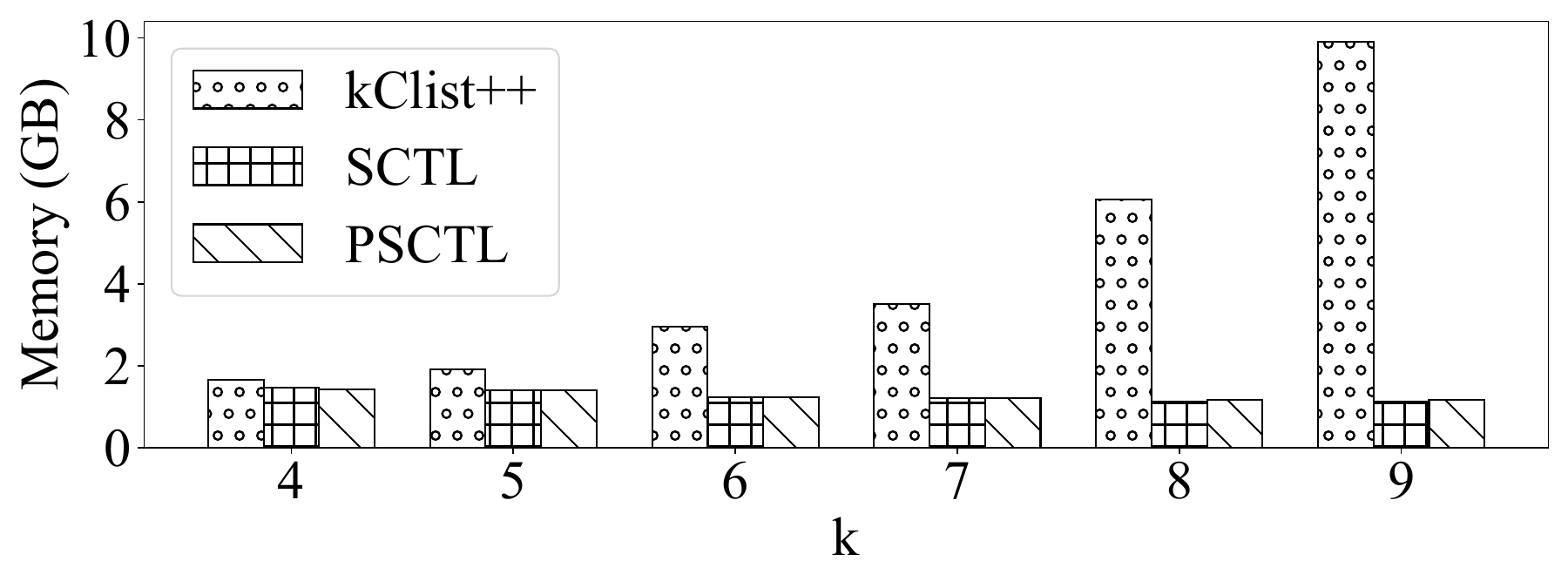}
		\end{center}
			\vspace*{-0.3cm}
		\caption{Memory costs of the Frank-Wolfe based algorithms on \pok.}
			\vspace*{-0.3cm}
		\label{fig:ffmem}
	\end{figure}

	\begin{figure*}[t]
		\begin{center}
			\begin{tabular}[t]{c}
				\subfigure[\epi]{\label{sfig:sta}
					\includegraphics[width=0.18\linewidth]{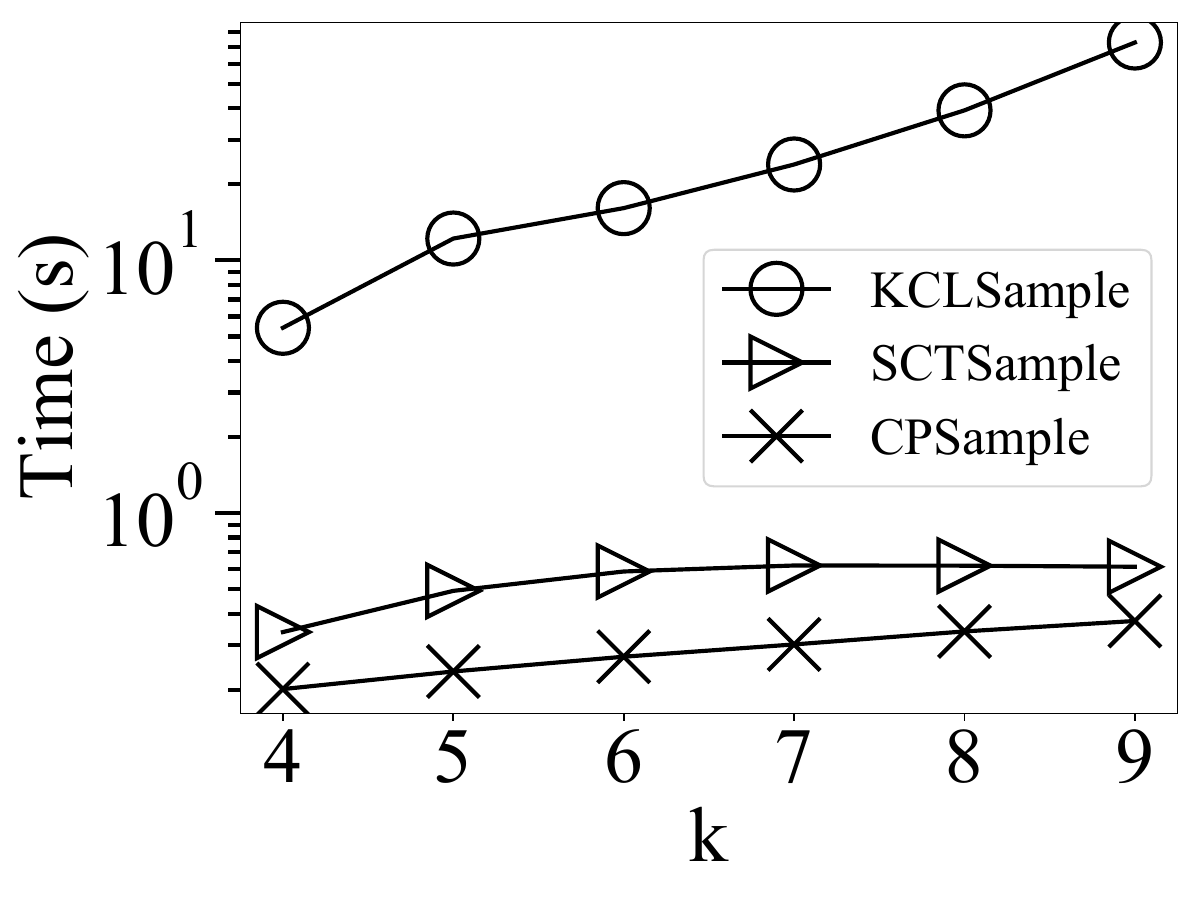}
				}
				\subfigure[\gow]{\label{sfig:stb}
					\includegraphics[width=0.18\linewidth]{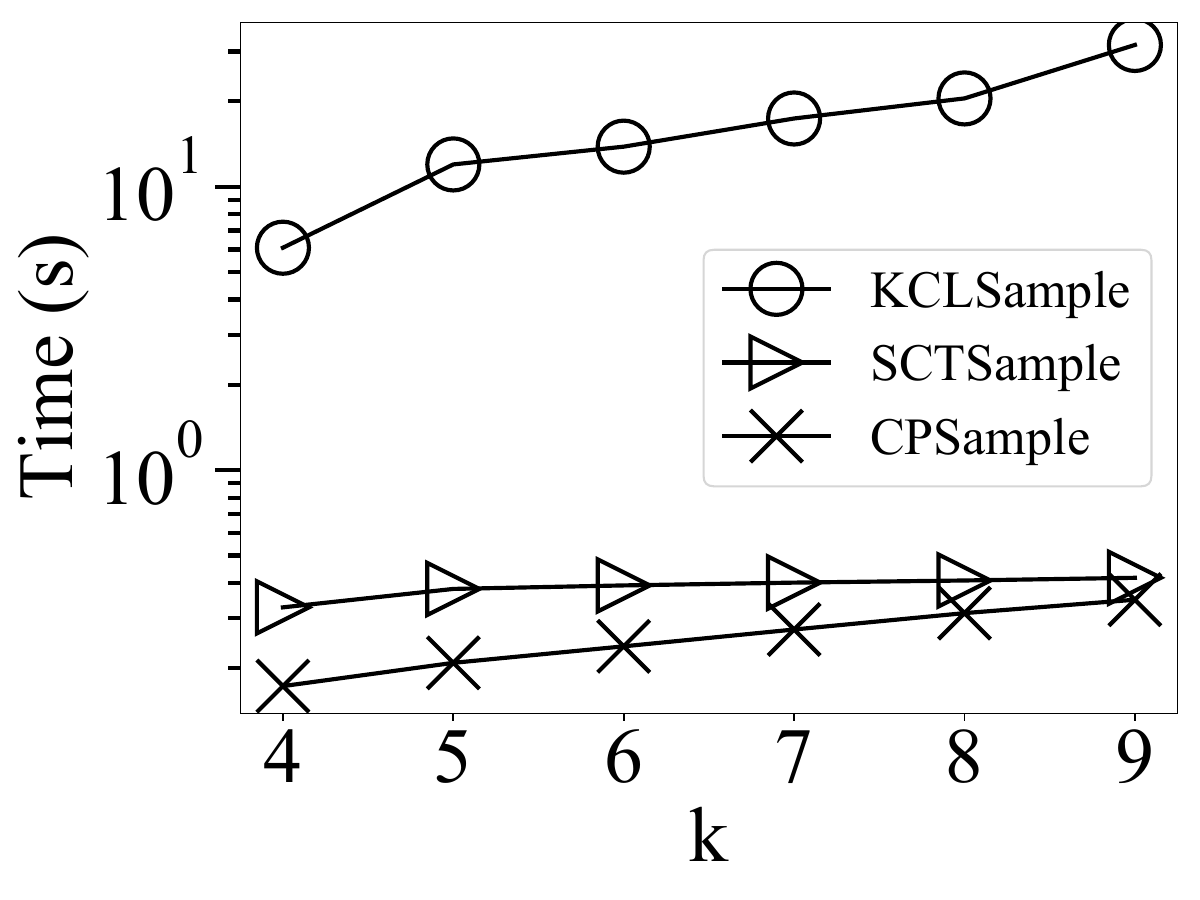}
				}
				\subfigure[\ama]{
					\includegraphics[width=0.18\linewidth]{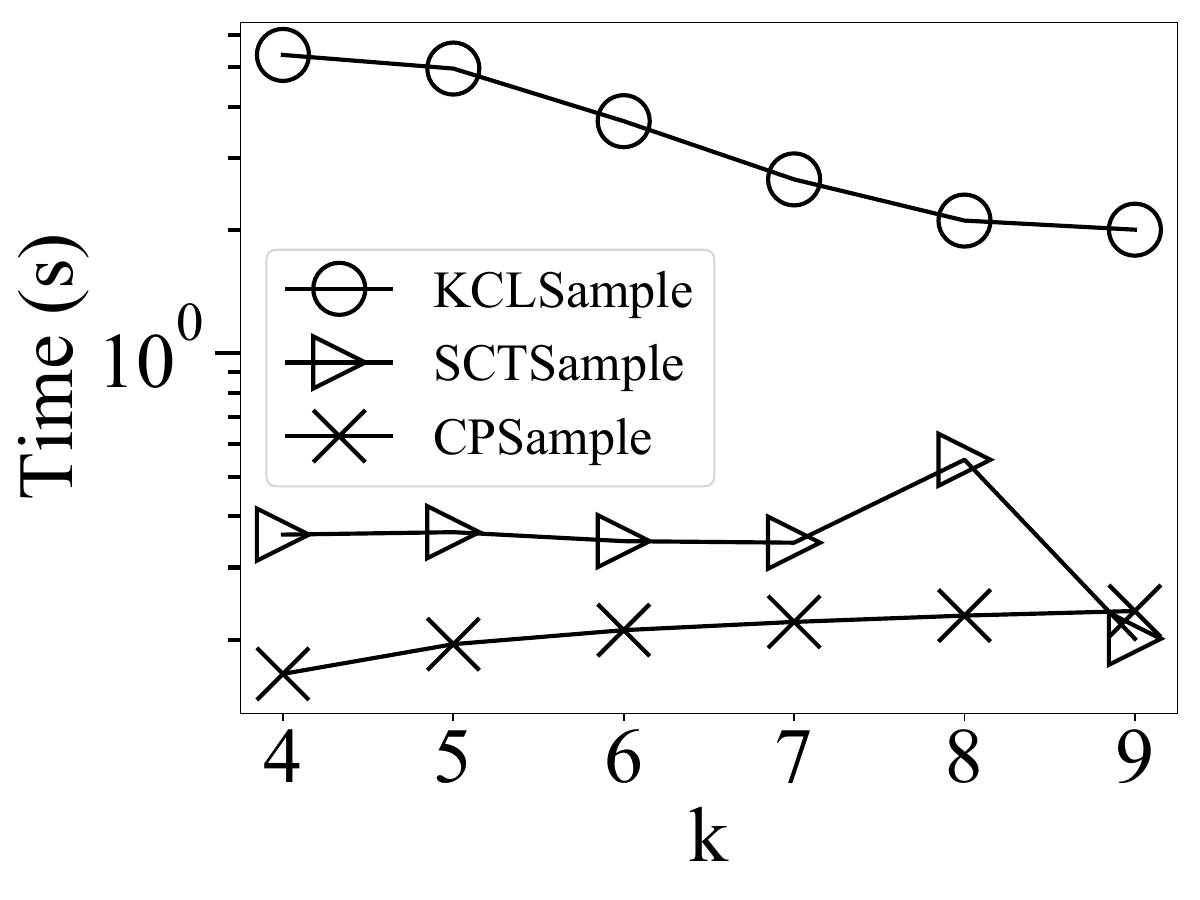}
				}
				\subfigure[\dbl]{\label{sfig:stc}
					\includegraphics[width=0.18\linewidth]{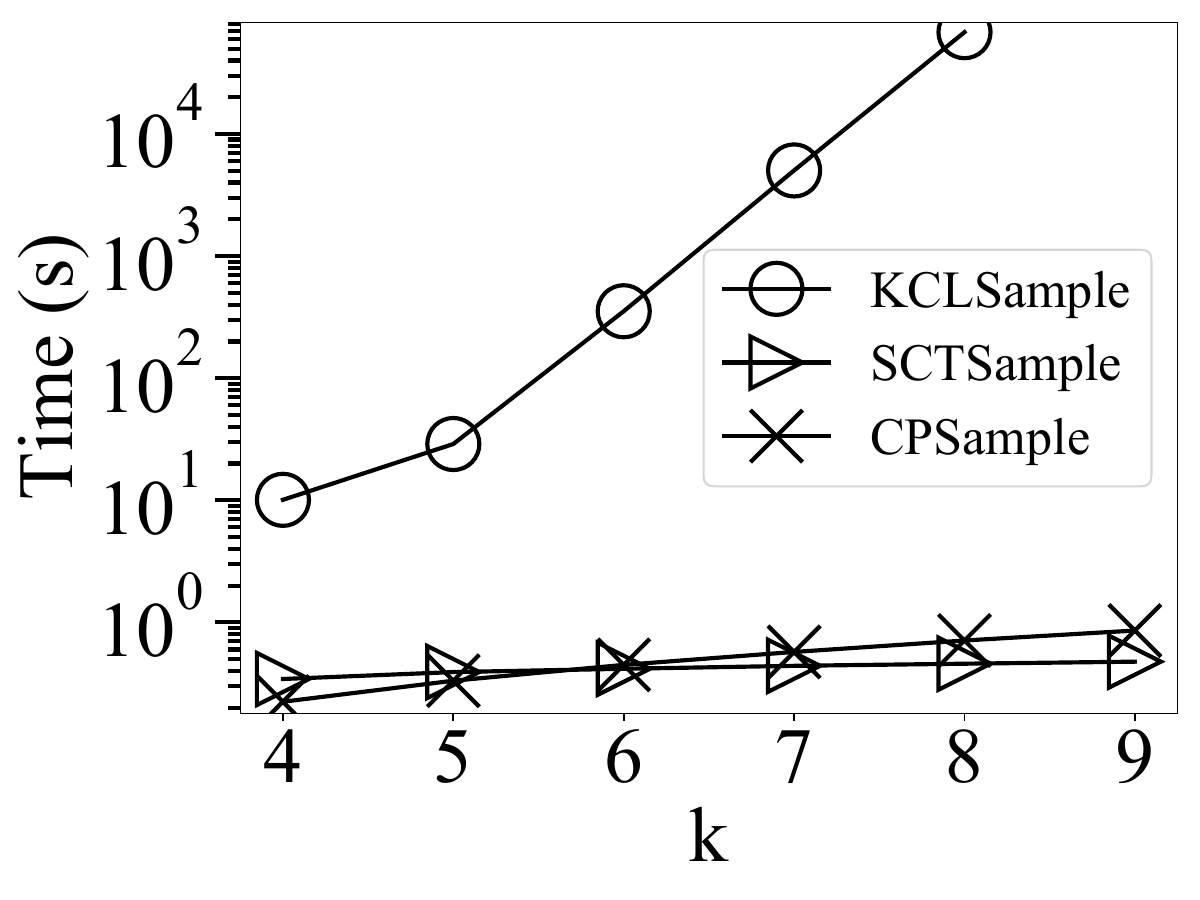}
				}
				\subfigure[\ber]{\label{sfig:std}
					\includegraphics[width=0.18\linewidth]{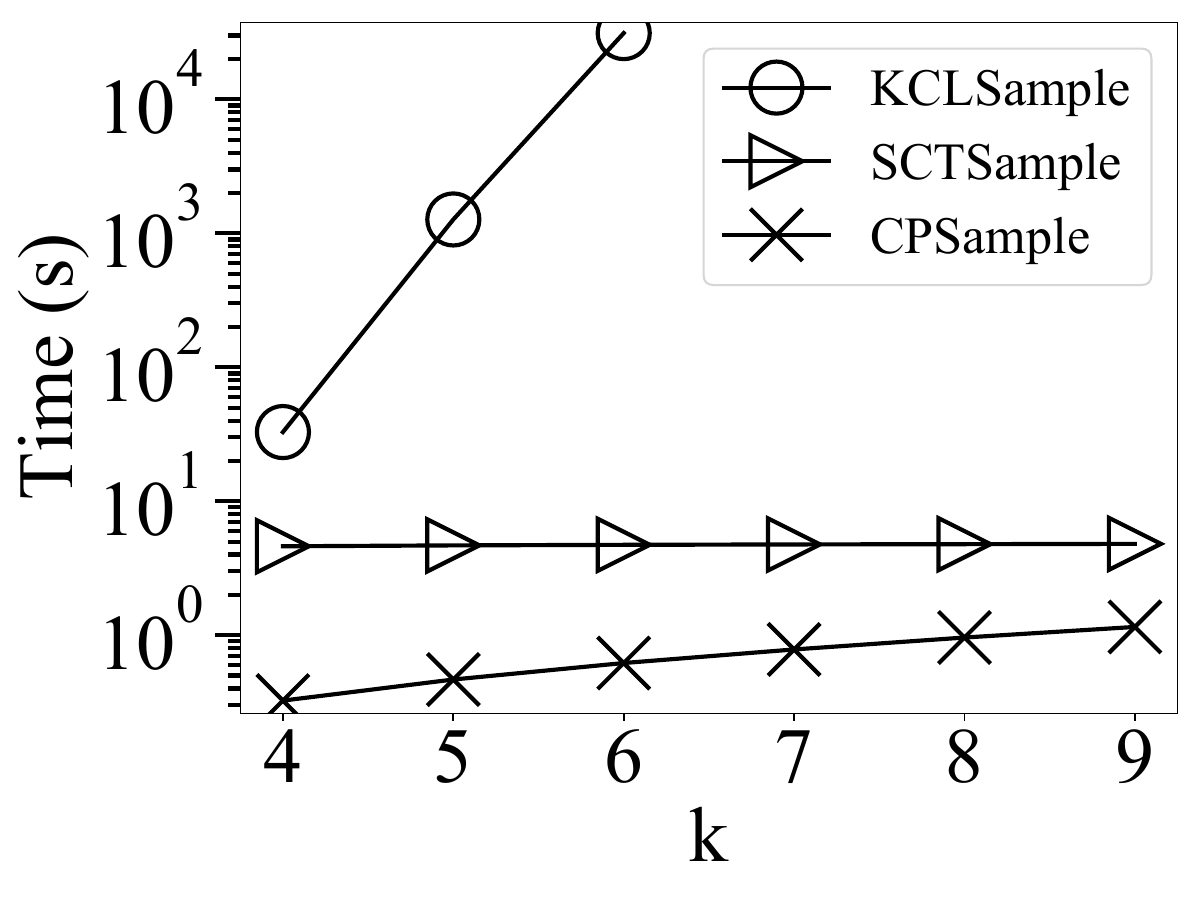}
				}
				\\
				\subfigure[\you]{
					\includegraphics[width=0.18\linewidth]{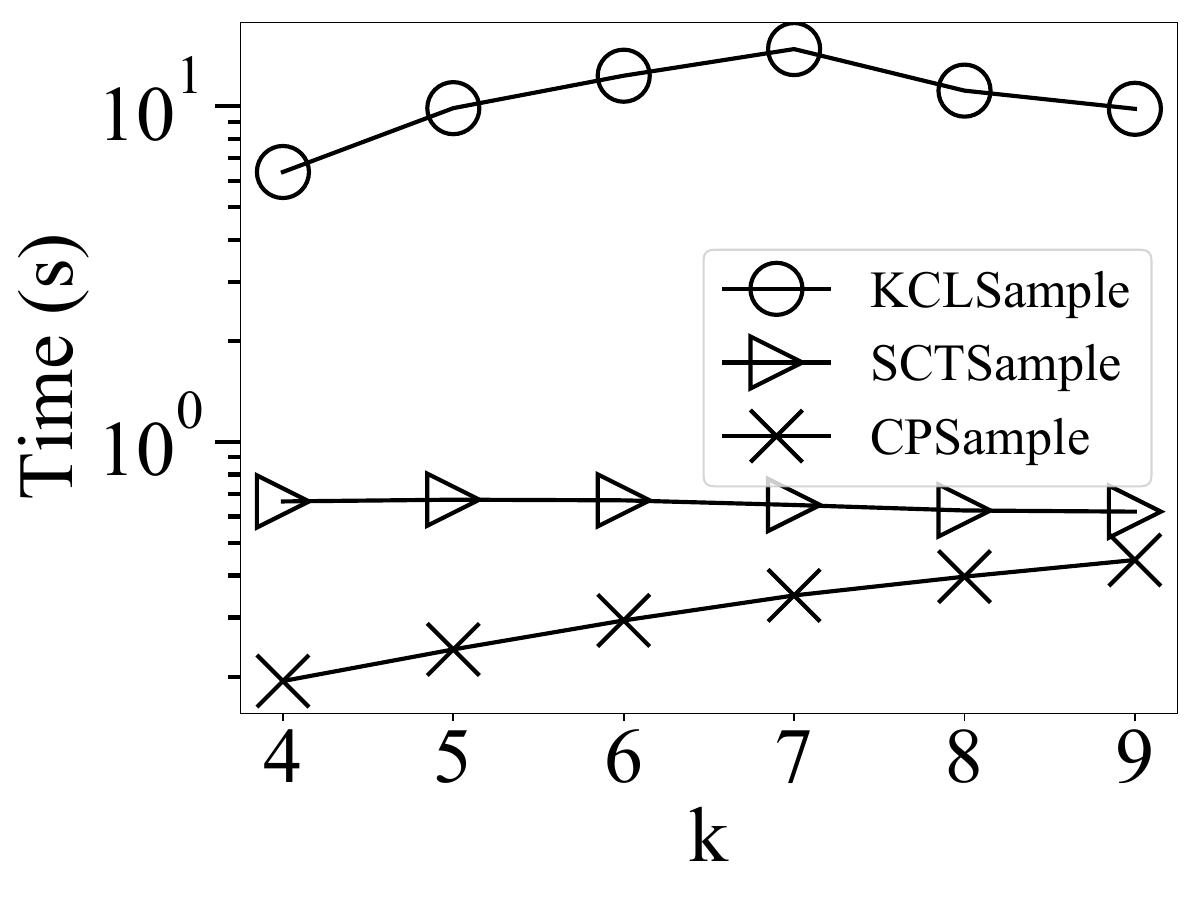}
				}
				
				\subfigure[\pok]{
					\includegraphics[width=0.18\linewidth]{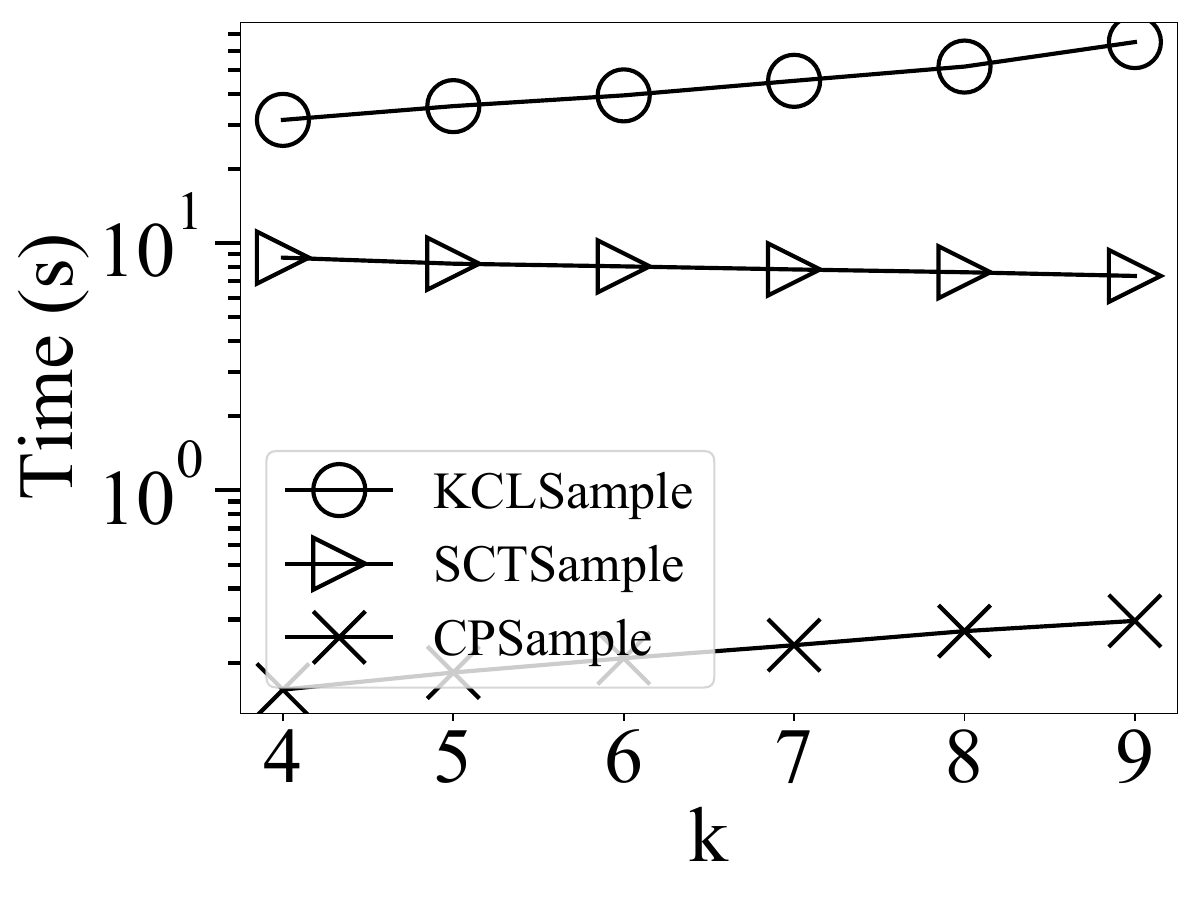}
				}
				\subfigure[\ski]{
					\includegraphics[width=0.18\linewidth]{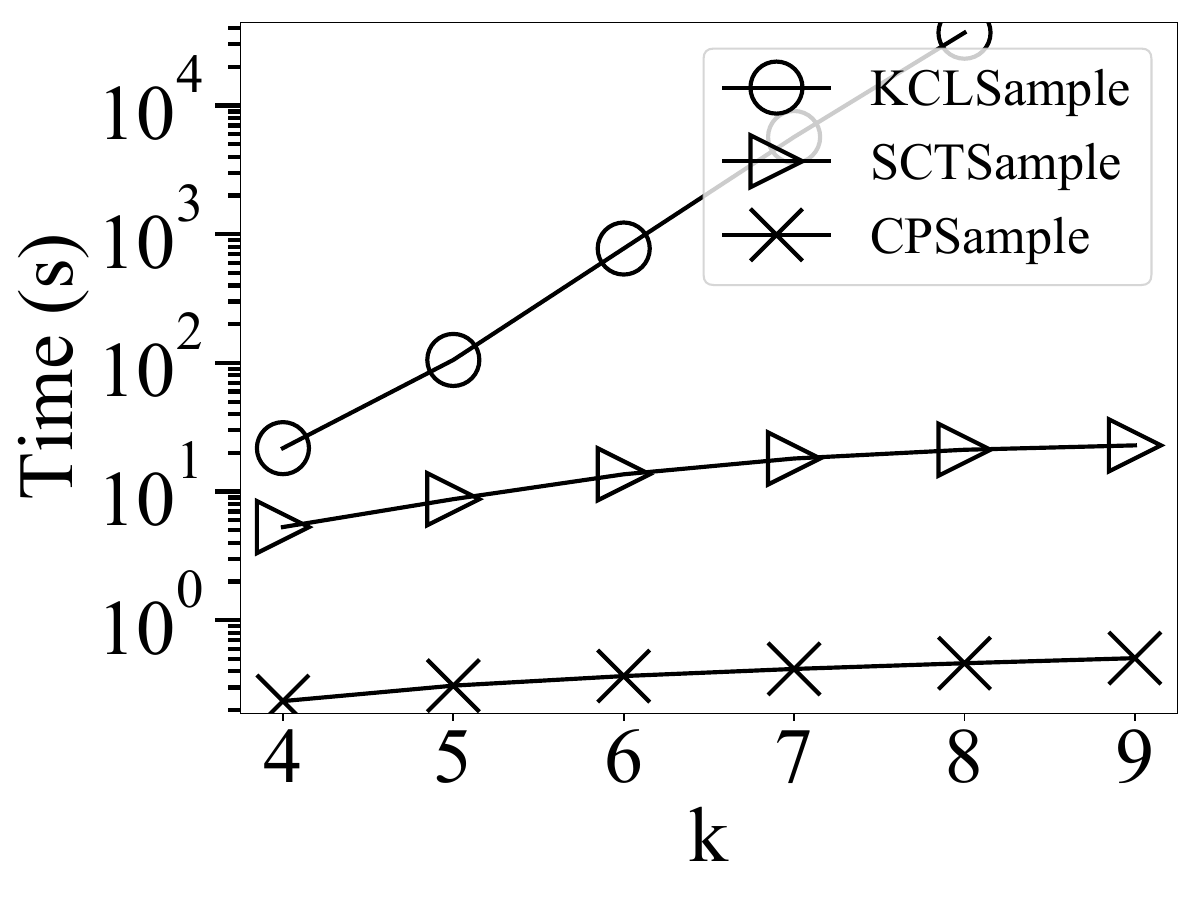}
				}
				\subfigure[\ork]{
					\includegraphics[width=0.18\linewidth]{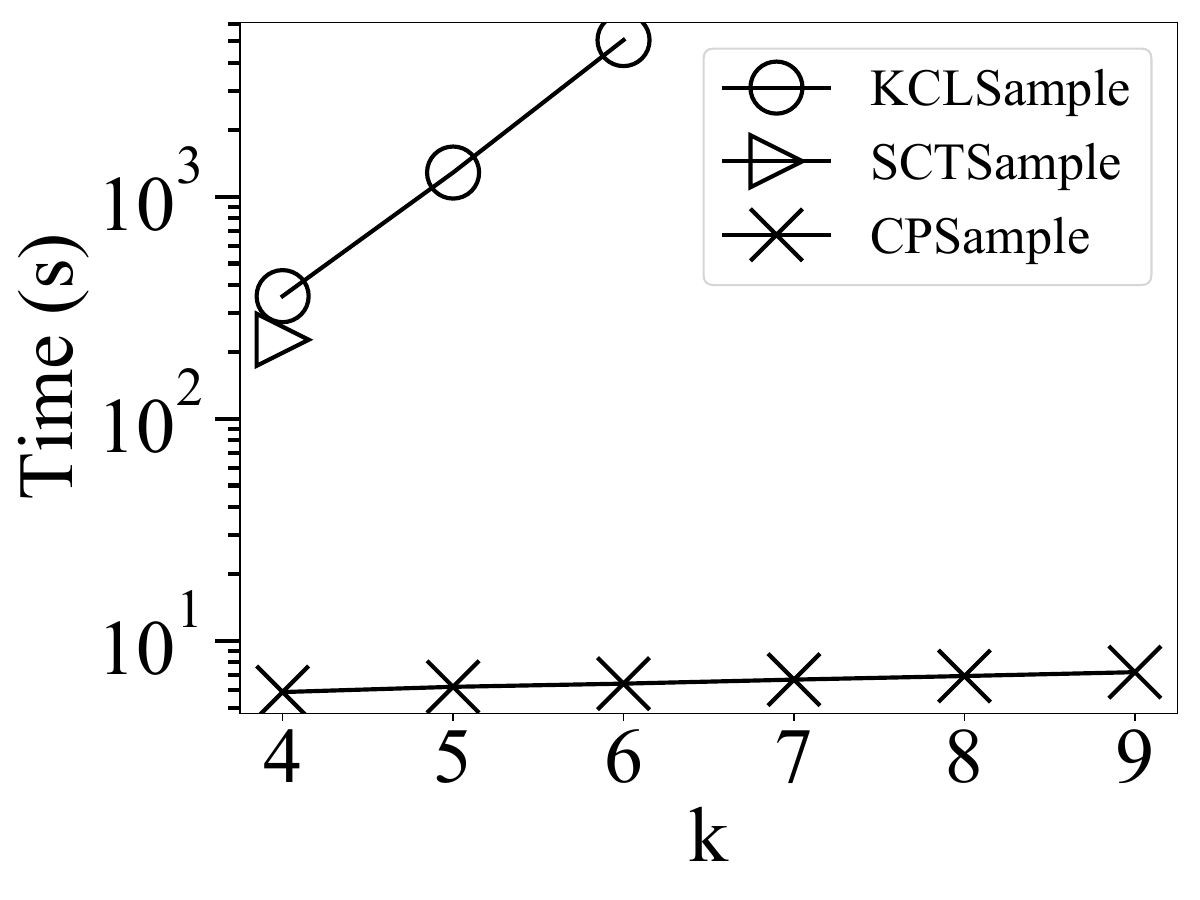}
				}
				\subfigure[\fri]{\label{sfig:stj}
					\includegraphics[width=0.18\linewidth]{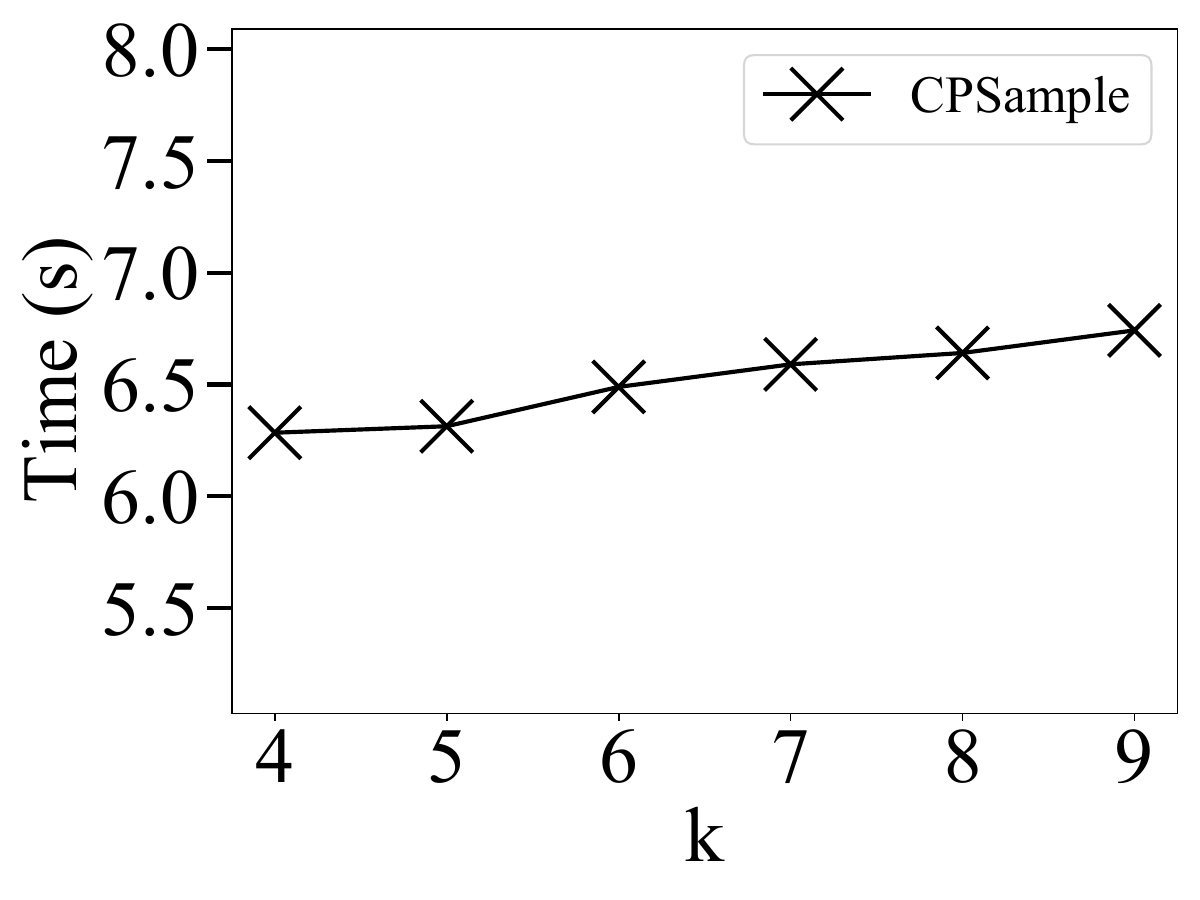}
				}
			\end{tabular}
		\end{center}
\vspace*{-0.2cm}
		\caption{Running time of the sampling based algorithms for various $k$. ($t=500000$).}
	\vspace*{-0.3cm}
		\label{fig:stime}
	\end{figure*}

	\begin{table*}[]
		\small
			\scriptsize
		\caption{The $k$-clique density obtained by different sampling algorithms within fixed time ($k=5$).}
			 \vspace*{-0.2cm} 	
		\label{tab:fixTime}
		\centering
			\begin{tabular}{c| c| c| c |c|c|c|c |c}
				\toprule
				\multirow{2}{*}{\textbf{Algorithms}} 
				&\multicolumn{2}{c|}{\dbl $\rho_k(V^*)=1287748.0$}
				& \multicolumn{2}{c|}{\pok $\rho_k(V^*)=11545.5$} & \multicolumn{2}{c|}{\ski $\rho_k(V^*)=1119664.6$}  & \multicolumn{2}{c}{\fri}  \\
				\cmidrule{2-9}
				&  \textbf{Time} & \textbf{$\rho_k(\hat{V^*})$} &
				\textbf{Time} & \textbf{${\rho_k(\hat{V^*})}$}  &  \textbf{Time} & \textbf{${\rho_k(\hat{V^*})}$} &   \textbf{Time} & \textbf{${\rho_k(\hat{V^*})}$}  \\

				\midrule
				\skcl & \multirow{3}{*}{$<0.1s$} & - 
				& \multirow{3}{*}{$<1s$} & - 
				&  \multirow{3}{*}{$<1s$} & -
				& \multirow{3}{*}{$<20s$} & - \\
				\ssctl &  & - & & - & & - & & -  \\
				\spath &  &  1287748.0 & & 11511.1 & & 1115421.7 & & 55815.2  \\
				\midrule
				\skcl & \multirow{3}{*}{$<1s$} & - 
				& \multirow{3}{*}{$<10s$} & - 
				& \multirow{3}{*}{$<10s$} & -
				& \multirow{3}{*}{$<60s$}&- \\
				\ssctl &  &  1287748.0& & 11545.5 & & 1119546.2& & - \\
				\spath & &  1287748.0 & & 11545.5 & & 1118411.5 & & 68573.8  \\
				\midrule
				\skcl & \multirow{3}{*}{$<20s$} &  1287748.0 
				& \multirow{3}{*}{$<30s$} &11545.5
				& \multirow{3}{*}{$<100s$} & 1119450.7 & \multirow{3}{*}{$<100s$}& - \\
				\ssctl &  & 1287748.0 & &11545.5 & & 1119664.6& & -  \\
				\spath & &  1287748.0 & &11545.5 & &1118808.0 & & 68748.8   \\
				
				\bottomrule
			\end{tabular}
	\end{table*}

\section{Experiments}

\stitle{Algorithms.} 
For the Frank-Wolfe based algorithms, we implement the \psctl algorithm which is based on Algorithm~\ref{alg:ibatch}. We use the state-of-the-art algorithm \kcl \cite{kclpp} and \sctl \cite{sctl} as the baseline algorithms.  \kcl and \sctl are all implementations of Algorithm~\ref{alg:framwork}. \kcl is a Frank-Wolfe based algorithm for \kcdsp that scan over the $k$-cliques individually through $k$-clique listing. \sctl is a Frank-Wolfe based algorithm, which scan over the $k$-cliques in batches through the SCT-index. 

For the sampling-based algorithms, we implement the \spath algorithm (Algorithm~\ref{alg:sample}). For comparison, we use the state-of-the-art sampling-based algorithms \skcl \cite{kclpp} and \ssctl \cite{sctl} as the baselines. Given a parameter $t$,  \kcl samples $t$ $k$-cliques during the procedure of $k$-clique listing, and \skcl samples $t$ $k$-cliques through SCT-index.

The source  code of \kcl and \skcl is  open available  \cite{kclpp}, which is implemented  in C++. Since the code of \sctl and \ssctl is not available, we implement them  by ourselves in C++, which shows similar performance compared to the results reported in \cite{sctl}.

\stitle{Datasets.} 
The details of the datasets are shown in Table~\ref{tab:networks}. The 5 columns of Table~\ref{tab:networks} are the dataset name, number of vertices, number of edges, degeneracy and the size of SCT respectively. 
All datasets are downloaded from \url{https://snap.stanford.edu/} or \url{http://konect.cc/}.

We evaluate all algorithms on a server with an AMD 3990X CPU and 256GB memory running Linux CentOS 7 operating system.

\subsection{Results of the FW-based algorihtms}

\stitle{Exp 1 : Runtime of various algorithms with varying $k$.} Figure~\ref{fig:fftime} shows the running time of \kcl, \sctl, and \psctl for varying $k$. To show the advantage of \ibatch, we omit the time taken by clique enumeration of \kcl and SCT construction. 

As $k$ increases, the running time of \psctl tends to be small. This is because the size of the SCT decreases, i.e. the value of $|\mathbb{P}|$ (Theorem~\ref{the:psctltime}) tends to be small as $k$ increases \cite{PIVOTER}. For example, on \pok, we have  $|\mathbb{P}|=10859743$ when $k=4$, while $|\mathbb{P}|=837568$ when $k=9$. As shown in Figure~\ref{fig:fftime}, \psctl substantially outperforms both \kcl and \sctl when $k=9$ on all the datasets. For example, on \dbl, our \psctl algorithm can achieve more than $5$ orders of magnitude faster than both \kcl and \sctl when $k \ge 7$. These results demonstrate the high efficiency of the proposed \psctl algorithm.

Additionally, on  the complex networks that the degeneracy is larger than $100$ (\dbl, \ber and \ski in Figure~\ref{fig:fftime}), our \psctl algorithm can consistently outperform the state-of-the-arts. On these networks, the count of cliques is very large. For example, \ber has $9.4\times 10^{12}$ $7$-cliques. The excellent performance of \psctl is due to the fact that the running time of \psctl is free from the count of $k$-cliques, which confirms our theoretical analysis in Section~\ref{sec:psctl}.

\stitle{Exp 2 : Runtime of various algorithms with varying $T$.}
Figure~\ref{fig:vart} shows the performance of the Frank-Wolfe based algorithms when $T$ varies  on Gowalla and Pokec. The results on the other datsets are consistent. We omit the time taken by clique enumeration and SCT construction. As expected, the running time of all algorithms is linear with respect to (w.r.t.)  $T$. Once again, our  \psctl is much more efficient  than the existing algorithms. These results further demonstrate the high efficiency of the proposed solution.

\stitle{Exp 3 : $k$-clique density with varying $T$.}
In Table~\ref{tab:convergence}, only \psctl can handle all the tested  networks.  \kcl can reach to the optimal results in only one iteration, but can not handle the complex networks with a large degeneracy, in which there exists a large number of $k$-cliques, like \ber. Both \sctl and \psctl can achieve a near-optimal approximation within a few iterations. These results further confirm the scalability of \psctl and the ability of \psctl to derive a good  approximation within only few iterations.

\stitle{Exp 4 : Running time needed to find $V^*$.} In the experiments, we find that \psctl can reach $V^*$ on all the tested datasets. Table~\ref{tab:convergence2} shows the number of iterations as well as the running time needed to find $V^*$. We find that \psctl achieves the lowest running time to obtain $V^*$ on $6$ datasets. Although \psctl requires $110$ iterations to converge to $V^*$ on \you, it can get a $99.999\%$ approximation using only $2.0$ seconds. These results further confirm the high efficiency of the proposed \psctl algorithm.

\stitle{Exp 5 : $k$-clique density within fixed time.} Figure~\ref{fig:fixTimeFW} compares the  $k$-clique density of the results of \sctl and \psctl with the same running time. In Figure~\ref{fig:fixTimeFW}, the results show that \psctl can get larger $k$-clique density than \sctl with the same running time. The results on other datasets is in Table~\ref{tab:convergence2} as described in Exp~4. 

\stitle{Exp 6 : Memory overheads.} We plot the memory costs of the Frank-Wolfe based algorithms in Figure~\ref{fig:ffmem}.  As can be seen, \psctl and \sctl have similar memory costs. This is because the memory costs of \sctl and \psctl are mainly taken by storing the SCT, while the memory cost of \kcl is  by  the storage of all $k$-cliques. As $k$ increases, the count of $k$-cliques grows and the size of SCT shrinks.These results indicate that our \psctl is space-efficient.

\subsection{ Results of the Sampling-based algorihtms}
\stitle{Exp 7 : Running time with varying $k$.}
In Figure~\ref{fig:stime}, we show the running time of the sampling-based algorithms for various values of $k$. From Figure~\ref{sfig:sta} to  \ref{sfig:stc}, we can see that  \spath and \ssctl achieve comparable performance. However, for large networks in Figure~\ref{sfig:std}$\sim$\ref{sfig:stj}, our \spath algorithm substantially outperforms \skcl and \ssctl. For example, on the \pok dataset, our algorithm is at least one order of magnitude faster than the two baselines. We also note that on the largest graph \fri, only our algorithm can obtain the results, while both of \skcl and \ssctl fail to drive the results. These results confirm the high efficiency and scalability of our sampling-based algorithm. 

\stitle{Exp 8 : $k$-clique density within fixed time.} In Table~\ref{tab:fixTime}, we compare $k$-clique density achieved by the sampling based algorithms within fixed running time.  As shown in Table~\ref{tab:fixTime}, our \spath one order of magnitude faster than existing algorithms and achieve good results. For example, on \dbl, \spath can terminate in $0.1s$ while \ssctl needs around $1s$ and \skcl needs about $20s$. Furthermore, our \spath is the only algorithm that can obtain the results on the largest graph \fri. These results confirm the high efficiency, effectiveness and scalability of our \spath.



\section{related work}

\stitle{Densest subgraph problem (DSP).} Given a graph and a measure of density, DSP requires to find a subset of vertices whose induced subgraph maximizes the value of density. DSP is a famous problem that has been studied for over five decades, which has a lot of variants and applications \cite{surveydensest}. For the traditional Edge-based Densest Subgraph Problem, several key algorithmic approaches have emerged.  These encompass  maximum-flow-based algorithms \cite{goldberg1984finding}, LP-based algorithms \cite{charikar2000greedy, Danisch17} and peeling-based algorithms \cite{charikar2000greedy, boob2020flowless}. There are also many variants of DSP. Densest $k$-subgraph problem aims at finding the densest subgraph with size $k$ \cite{Feige2001Dense}; Densest at-least(most)-k-subgraph problem aims at finding the densest subgraph with size at least(most) $k$ \cite{Andersen2009Finding}; Anchored Densest Subgraph Problem tries to find the densest subgraph that contains a given seed set \cite{Dai2022Anchored}; Fair densest subgraph is the densest subgraphs that has equal represented colors \cite{DBLP:conf/cikm/Anagnostopoulos20, DBLP:conf/kdd/0001CST23}; Motif-based densest subgraph is the generalized version of $k$-clique densest subgraph, where the density is defined by the count of a given motif \cite{FangYCLL19Efficient}. On different graphs like directed graphs \cite{charikar2000greedy, ma2020efficient},  temporal graphs \cite{BogdanovMS11Mining} and hypergraphs \cite{HuangK95}, there also exists the corresponding DSPs.

\stitle{Large near-clique detection.}The maximum clique problem is an important problem which has a lot of applications \cite{Tomita17, Chang20MC}. However, maximum clique often has a very tight constraints for real-world applications.  Thus, a lot of relaxed models for large near-clique detection were proposed \cite{kdensWWW2015,VladimirCore,BalasundaramBH11,maximumDclique21}. The $k$-clique densest subgraph studied in this work is known as a large near-clique. State-of-the-art algorithms for solving \kcdsp are primarily rooted in the Frank-Wolfe framework, because \kcdsp can be formulated as a convex programming. Sun \cite{kclpp} introduced the first Frank-Wolfe-based algorithm, named \kcl.  \kcl operates by sequentially iterating over individual $k$-cliques, and each $k$-clique assigns a weight to the vertex with the smallest weight among the $k$ vertices. With an adequate number of iterations, it can converge to the optimal solution $V^*$. Recent advancements by He et al. \cite{sctl} have accelerated \kcl, \kclexact and \skcl by the technique SCT-index. The SCT-index, building upon the SCT-tree data structure \cite{PIVOTER}, enables batch-wise iteration over $k$-cliques, significantly enhancing efficiency. Other methods include flow-based algorithms \cite{MitzenmacherPPT15, kdensWWW2015} and peeling-based algorithms \cite{FangYCLL19Efficient}, but they are not as efficient as the Frank-Wolfe based algorithms \cite{sctl}. There also exists a lot of other near-clique models \cite{VladimirCore, BalasundaramBH11, ChangDef}. Maximum $k$-core is the largest subgraph that each vertex has degree larger than $k$ \cite{VladimirCore}. Maximum $k$-plex is the maximum subgraph that each vertex has at most $k$ non-neighbors \cite{BalasundaramBH11, ChangXS22}. Maximum $s$-defective clique is the maximum subgraph that misses at most $s$ edges compared to clique \cite{maximumDclique21, ChangDef, dcliqueAAAI2022}.

\section{conclusion}
In this paper, we study the problem of $k$-clique densest subgraph search. We propose a new Frank-Wolfe-based algorithm, whose time complexity is free from the count of $k$-cliques. Thus, it is very efficient for processing large graphs that often have a extremely number of $k$-cliques. We present a detailed theoretical analysis of our algorithms. To further improve the efficiency, we also propose a new and provable sampling-based algorithm. A nice feature of our algorithm is that it has polynomial time complexity. We conduct extensive experiments on 12 large real-world graphs to evaluate our algorithms, and the results demonstrate the high efficiency and scalability of our approaches.

\bibliographystyle{ACM-Reference-Format}
\balance
\bibliography{reference}

\end{document}